%% file: main.tex
\documentclass[10pt,twocolumn,letterpaper]{article}

\usepackage{iccv}  
\usepackage[accsupp]{axessibility} 
\usepackage{amsmath,amssymb,amsfonts}
\usepackage{algorithmic}
\usepackage{graphicx}
\usepackage{textcomp}
\usepackage{xcolor}
\usepackage{booktabs}
\usepackage{multirow}
\usepackage{array}
\usepackage{float}
\usepackage{breakcites}


\definecolor{iccvblue}{rgb}{0.21,0.49,0.74}
\usepackage[pagebackref,breaklinks,colorlinks,allcolors=iccvblue]{hyperref}

\def\paperID{15060} 
\def\confName{ICCV}
\def\confYear{2025}

\title{MH-LVC: Multi-Hypothesis Temporal Prediction for Learned Conditional Residual Video Coding}

\author{Huu-Tai Phung\textsuperscript{1}\footnotemark[1]
\qquad Zong-Lin Gao\textsuperscript{1}\footnotemark[1]
\qquad Yi-Chen Yao\textsuperscript{1}
\qquad Kuan-Wei Ho\textsuperscript{1}
\qquad Yi-Hsin Chen\textsuperscript{1} \\
\qquad Yu-Hsiang Lin\textsuperscript{1} 
\qquad Alessandro Gnutti\textsuperscript{2} 
\qquad Wen-Hsiao Peng\textsuperscript{1} \\\\
\textsuperscript{1} National Yang Ming Chiao Tung University, Taiwan \\
\textsuperscript{2}	University of Brescia, Italy \\ 
}

\newcommand{\beginsupplement}{%
        \setcounter{table}{0}
        \renewcommand{\thetable}{A\arabic{table}}%
        \setcounter{figure}{0}
        \renewcommand{\thefigure}{A\arabic{figure}}%
        \setcounter{section}{0}
        \renewcommand{\thesection}{A\arabic{section}}%

        \setcounter{footnote}{0}
        \renewcommand{\thefootnote}{\fnsymbol{footnote}}
}

\begin{document}
\maketitle
\renewcommand*{\thefootnote}{\fnsymbol{footnote}}
\footnotetext[1]{Equal contribution.}
\input{section/0_Abstract}
\input{section/1_Introduction}
\input{section/2_related_work}

\input{section/3_Method}
\input{section/4_Result}
\input{section/5_Conclusion}

\section*{Acknowledgement}
\noindent This work is supported by MediaTek and National Science and Technology Council (NSTC), Taiwan, under Grants 113-2634-F-A49-007- and 111-2923-E-A49-007-MY3. We thank National Center for High-performance Computing (NCHC) for providing computational and storage resources, and NVAITC for providing access to the Taipei-1 supercomputer.

{
    \small
    \bibliographystyle{ieeenat_fullname}
    \bibliography{main}
}

\input{supp}

\end{document}

%% file: section/0_Abstract.tex
\begin{abstract}

This work, termed MH-LVC, presents a multi-hypothesis temporal prediction scheme that employs long- and short-term reference frames in a conditional residual video coding framework. Recent temporal context mining approaches to conditional video coding offer superior coding performance. However, the need to store and access a large amount of implicit contextual information extracted from past decoded frames in decoding a video frame poses a challenge due to excessive memory access. Our MH-LVC overcomes this issue by storing multiple long- and short-term reference frames but limiting the number of reference frames used at a time for temporal prediction to two. Our decoded frame buffer management allows the encoder to flexibly utilize the long-term key frames to mitigate temporal cascading errors and the short-term reference frames to minimize prediction errors. Moreover, our buffering scheme enables the temporal prediction structure to be adapted to individual input videos. While this flexibility is common in traditional video codecs, it has not been fully explored for learned video codecs. Extensive experiments show that the proposed method outperforms VTM-17.0 under the low-delay B configuration in terms of PSNR-RGB across commonly used test datasets, and performs comparably to the state-of-the-art learned codecs (e.g.~DCVC-FM) while requiring less decoded frame buffer and similar decoding time. 


\end{abstract}
\vspace{-1 em}

%% file: section/1_Introduction.tex
\section{Introduction}
\label{sec:intro}

\input{section/1-1_Teaser}

How to leverage the information from the past decoded frames has been the central theme of video coding, learned and traditional, with uni-directional temporal prediction. Similar to traditional codecs, some early learned codecs~\cite{mlvc,vlvc} explicitly store multiple decoded frames in the decoded frame buffer. However, unlike traditional codecs, they fuse these reference frames simultaneously with a network~\cite{mlvc} or a weighted trilinear warping~\cite{vlvc} to construct a high-quality temporal predictor. Many state-of-the-art learned codecs~\cite{tcmli, hemli, dcvcdc, dcvcfm, RLVC} adopt a data-driven approach, using a recurrent neural network-like mechanism to learn contextual information from past decoded frames. This method, known as temporal context mining, requires a large buffer to retain and propagate implicit information along the temporal dimension (see Fig.~\ref{fig:Teaser}~(a)). For example, DCVC-FM~\cite{dcvcfm} stores 48+ feature maps $F_t$ with the same spatial resolution as that of the input video frame to decode a single video frame.

Recent works~\cite{dcvcpqa, dcvcsdd} along this line of research even double the buffering requirements to explore the full potential of conditional video coding. Because of the sheer amount of information needed to be stored, these high-resolution latent features may need to be kept in the external (off-chip) memory when it comes to their ASIC implementation. Fetching high-resolution features/frames in encoding/decoding a single frame imposes a heavy burden on memory access. This issue becomes more prominent when encoding/decoding high frame-rate, high-resolution videos (e.g. 4K videos at 60 frames per second).

To address this issue, we propose a multi-hypothesis temporal prediction scheme in a conditional residual coding framework, termed MH-LVC, for learned video coding. We draw inspiration from traditional codecs to avoid excessive access to the decoded frame buffer. Specifically, in encoding/decoding a video frame, the number of reference frames used at a time for temporal prediction is limited to two. This constraint allows multiple reference frames to be stored and accessed for temporal prediction without significantly increasing the memory access bandwidth between the processing units and the decoded frame buffer. Notably, one of the reference frames is chosen from the decoded frames with long prediction distances, referred to as the long-term key frames, to mitigate temporal cascading errors (i.e. the degradation of the decoded image quality over time). The other reference frame is the most recently decoded frame, chosen for its short prediction distance with lower prediction errors. The long- and short-term reference frames complement each other via a spatially adaptive fusion in formulating a temporal predictor. 

As illustrated in Fig.~\ref{fig:Teaser} (b) and summarized as follows, the novelty of this work includes: (1) a novel attempt to leverage the long- and short-term temporal prediction in a way that avoids excessive access to the decoded frame buffer while exploiting effectively the contextual information from the past to improve coding efficiency, (2) a spatially adaptive temporal prediction module that fuses long- and short-term reference frames at the sub-frame level to mitigate prediction errors caused by long prediction distances, and (3) an unexplored attempt to adapt temporal prediction structure to input videos during inference. Although the explicit buffer management has been widely adopted by traditional video codecs, to our best knowledge, it has not been explored for learned video codecs.

For experiments, we implemented MH-LVC based on the same masked conditional residual coding framework in~\cite{maskcrt}, but replaced its Transformer-based backbone with the CNN-based one adapted from DCVC-FM~\cite{dcvcfm}. The intensive results confirm that our MH-LVC outperforms VTM-17.0 under the low-delay B configuration on HEVC-B, UVG, and MCL-JCV datasets in terms of PSNR-RGB.
It performs similarly to DCVC-FM on most test datasets, with a much smaller decoded frame buffer and a similar decoding time. In some settings, our MH-LVC has a much higher encoding runtime than DCVC-FM due to the online adaptation of the prediction structure. This encoder issue will be further addressed by developing a fast frame-selection algorithm in the future.

%% file: section/1-1_Teaser.tex
\begin{figure}[t]
    \centering
    
    \begin{subfigure}[]{0.47\textwidth}
        \label{fig:Teaser-a}
        \centering
        \includegraphics[width=\textwidth, trim= 0 2500 0 0, clip]{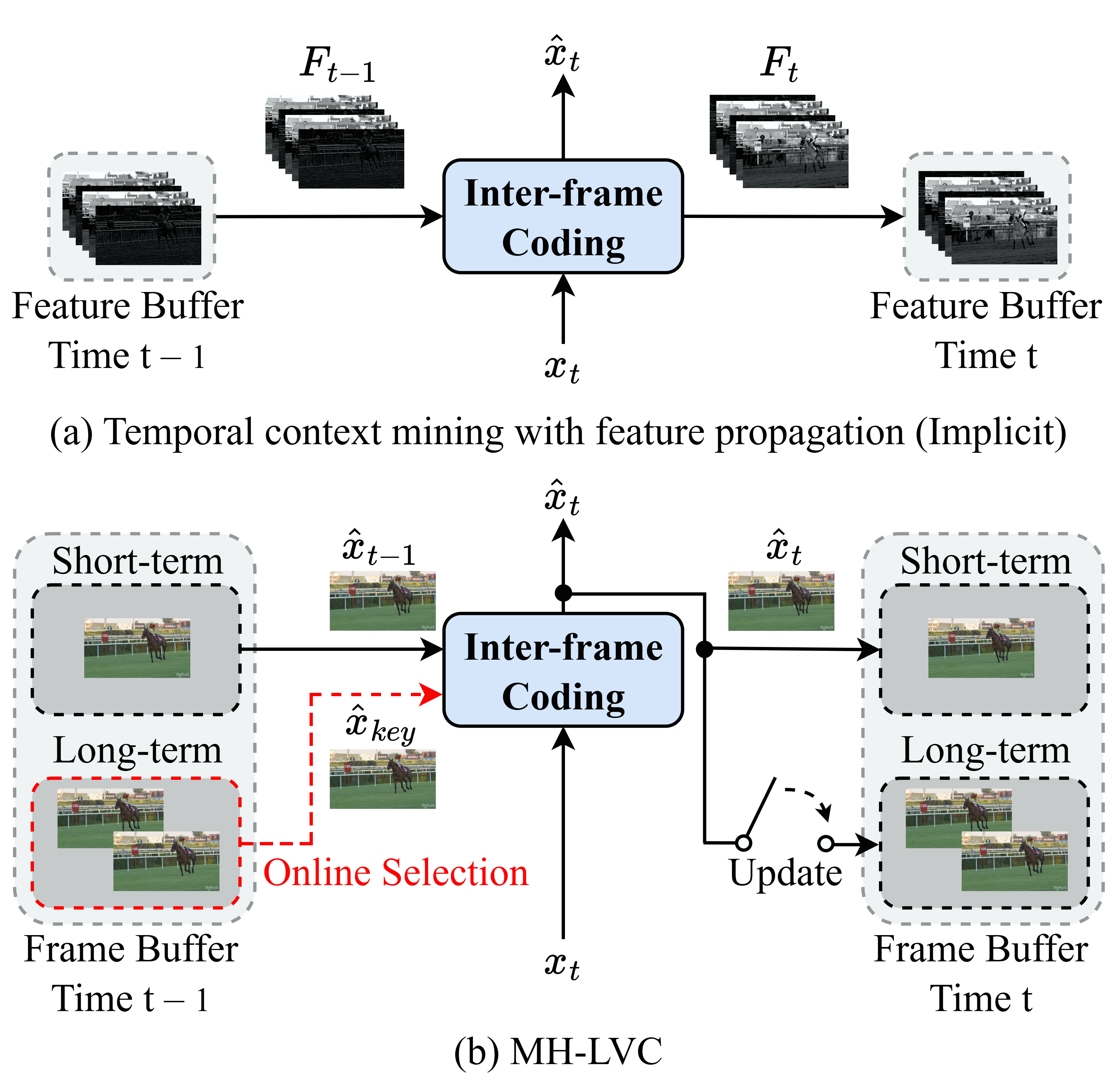}
        \caption{Temporal context mining with feature propagation}
    \end{subfigure}
    
    \begin{subfigure}[]{0.47\textwidth}
        \label{fig:Teaser-b}
        \centering
        \includegraphics[width=\textwidth, trim= 0 220 0 1700, clip]{figure/Introduction/teaser_ICCV.png}
        \caption{MH-LVC}
    \end{subfigure} 

    \caption{Comparison of how temporal contextual information is formulated for inter-frame coding. (a) The temporal context mining learns latent features $F_{t-1}$ to be stored, propagated, and updated temporally as $F_t$ in coding the input frame $x_t$. (b) Our MH-LVC maintains multiple previously decoded frames as long- and short-term reference frames, with only one short-term frame $\hat{x}_{t-1}$ and one long-term key frame $\hat{x}_{key}$ selected adaptively for coding $x_t$.}
   
    \label{fig:Teaser}
    \vspace{-1 em}
\end{figure}

%% file: section/2_related_work.tex
\section{Related Work}
\label{sec:Related_work}

Temporal prediction plays a key role in end-to-end learned (and traditional) video codecs. Most learned codecs focus on uni-directional temporal prediction, with the aim to exploit the decoded frames/features in the past for better coding an input frame. In the residual coding framework~\cite{dvclu, fvc, mlvc, elfvc}, these decoded frames/features are used to construct a temporal predictor with motion compensation to generate a residual frame for coding. In the more advanced conditional coding framework~\cite{dcvc, canfvc, dcvcfm}, they are used to generate condition signals to condition the autoencoder for encoding/decoding the input frame. The emerging conditional residual coding scheme~\cite{cond_res_coding} hybrids residual and conditional coding. The contextual information from the past is used to formulate a temporal predictor to predict the input frame, and to construct the condition signals for coding the residual frame resulting from the prediction. More recently, masked conditional residual coding (MaskCRT)~\cite{maskcrt} hybrids conditional coding and conditional residual coding at the sub-frame level.

The way the contextual information is aggregated from the past decoded frames has a significant impact on coding efficiency and complexity. One common approach~\cite{dvclu,ssf,dmvc,maskcrt} is to reference the previously decoded frame in formulating the temporal predictor and/or condition signals, an idea similar to uni-directional, single-hypothesis temporal prediction in traditional codecs. It is enabled by allocating a decoded frame buffer to store the short-term reference frame. The works in~\cite{mlvc,vlvc} buffer more short-term reference frames for multi-hypothesis prediction at the cost of increased buffer size and more frequent access to the decoded frame buffer. In a sense, these codecs implement a recurrent neural network with only output recurrence (i.e. decoded frames as the only contextual information from the past) without maintaining and propagating latent states temporally. This is in direct contrast to another school of thought~\cite{RLVC, tcmli, hemli, dcvcdc, dcvcmip, dcvcfm, dcvclcg} known as temporal context mining. It adopts a data-driven approach to learn and propagate latent features in addition to decoded frames. This implicit buffering scheme amounts to constructing a recurrent neural network with both output recurrence and hidden connections, the design of which is theoretically more powerful than the network with only output recurrence. Due to their learning-based nature, there is little control of buffered contents. This makes it challenging to perform online adaptation based on the characteristics of the input video. Moreover, these approaches usually store a large number of latent features, which must be accessed simultaneously in encoding/decoding a video frame, causing the issue of excessive memory access bandwidth.

%% file: section/3_Method.tex

\section{Proposed Method}
\label{sec:Proposed_Method}
\input{section/3-1_OverView}

This work (MH-LVC) introduces a multi-hypothesis temporal prediction scheme in a conditional residual coding framework. Unlike the existing works with multiple reference frames, it has the striking feature of limiting the number of temporal prediction hypotheses used at a time to two in order to strike a balance between coding performance and memory access bandwidth. A similar constraint has been adopted by the standards community in developing traditional video codecs, in order to limit the memory access bandwidth. In our work, one of the reference frames is adaptively chosen from previously decoded frames with long prediction distances, and this selection is signaled in the bitstream for video frame decoding. The other frame is the most recently decoded frame for encoding complexity considerations. Unlike implicit temporal context mining techniques~\cite{tcmli, hemli, dcvcdc, dcvcfm, RLVC}, MH-LVC allows the temporal prediction structure to be either specified manually or adapted online to input videos. 



\subsection{System Overview}
\label{sec:sys_overview} 
Fig.~\ref{fig:OverView} illustrates our MH-LVC framework. It includes three main components: a motion codec $\{M^{enc}, M^{dec}\}$, an inter-frame codec $\{G^{enc}, G^{dec}\}$, and a long- and short-term temporal prediction (LSTP) module. Both the motion and inter-frame codecs are built on the same conditional residual coding framework in~\cite{maskcrt}, but with its Transformer-based backbone replaced with the CNN-based backbone adapted from DCVC-FM~\cite{dcvcfm}.

Encoding a video frame $x_t \in \mathbb{R}^{3 \times W \times H}$ of width $W$ and height $H$ begins by estimating an optical flow map $f_t \in \mathbb{R}^{2 \times W \times H}$ with respect to its decoded reference frame $\hat{x}_{t-1}$. The flow map $f_t$ is coded by the motion codec $\{M^{enc}, M^{dec}\}$, the process of which requires the previously decoded flow latents $\hat{y}^{f}_{t-1}$ and the flow features $F^{f}_{t-1}$ extracted from the motion decoding process.

For inter-frame coding, we adaptively fuse two reference frames chosen from the decoded frame buffer to formulate the temporal predictor $x_c \in \mathbb{R}^{3 \times H \times W}$ and the multi-scale condition signals $C^1 \in \mathbb{R}^{48 \times H \times W},~C^2 \in \mathbb{R}^{64 \times H/2 \times W/2},~C^3 \in \mathbb{R}^{96 \times H/4 \times W/4}$. Specifically, the most recently decoded frame $\hat{x}_{t-1}$, i.e. a short-term reference frame, and its decoded flow map $\hat{f}_{t}$ are used as the first temporal hypothesis. Additionally, a second hypothesis is created from a long-term key frame $\hat{x}_{key}$ along with a long-term flow map $\hat{f}^{acc}_{t}$, which is derived from the decoded short-term flow maps that describe the motion of consecutive frames that sit between $\hat{x}_{key}$ and $x_t$. Notably, $\hat{x}_{key}$ is allowed to be chosen during inference according to a pre-defined temporal prediction structure or adaptively based on a rate-distortion criterion. To this end, the decoded frame buffer stores multiple reference frames with a buffer management scheme. Although there are multiple reference frames to choose from, we specifically limit the number of reference frames fused at a time to two. This is done to avoid excessive access to the decoded frame buffer, which often resides in the external memory. 

Our LSTP module fuses $\hat{x}_{t-1},\hat{x}_{key}$ in a spatially adaptive manner through a gating mechanism. Conceptually, it performs adaptive fusion at the sub-frame level. This is distinct from the early attempt to incorporate multiple reference frames~\cite{mlvc} into learned codecs, where all the reference frames in the decoded frame buffer are fused without considering the reliability of these reference frames. 

Lastly, as a conditional residual coding scheme, our MH-LVC follows~\cite{maskcrt} in generating the input to $G^{enc}$ and reconstructing it from $G^{dec}$, using a soft mask $m \in \mathbb{R}^{3 \times H \times W}$ predicted from $\hat{f}_t$ and $x_c$ to weight $x_c$ for conditional residual coding. 


\subsection{Decoded Frame Buffer Management}
\label{ssec:Periodic_refresf}

The decoded frame buffer stores previously decoded frames and flow maps to implement the uni-directional temporal prediction with multiple hypotheses. It includes a short-term section and a long-term section. The short-term section stores the most recently decoded frame $\hat{x}_{t-1}$ as the short-term reference, along with the accompanying decoded flow map $\hat{f}_t$ for temporal prediction. The long-term section keeps multiple long-term key frames and their respective accumulated flow maps, which are updated on-the-fly according to Eq.~\eqref{eq:acc} to keep track of the motion between a coding frame $x_t$ and these long-term key frames. 
\begin{equation}
\label{eq:acc}
\hat{f}^{acc}_{new} = \hat{f}_t +Warp(\hat{f}^{acc}_{old}, \hat{f}_t)
\end{equation} 
where $\hat{f}^{acc}_{old},\hat{f}^{acc}_{new}$ represent respectively the accumulated long-term flow maps before and after being updated with the currently decoded flow map $\hat{f}_t$. $Warp(\hat{f}^{acc}_{old}, \hat{f}_t)$ denotes the backward warping operation, which warps $\hat{f}^{acc}_{old}$ by $\hat{f}_t$.

Fig.~\ref{fig:CondingStructure} depicts how the decoded frame buffer evolves over time for implementing our current temporal prediction structure. The two latest long-term key frames are buffered in first-in-first-out order. To mitigate temporal cascading errors, this work introduces a 4-frame mini-group-of-pictures (mini-GOP) and a nested frame quality structure. The last frames in mini-GOPs are marked periodically as long-term key frames and coded at the highest quality level to facilitate temporal prediction of video frames in the subsequent mini-GOPs. Notably, this mechanism allows the most recently decoded frame to be included in the long-term section. As mentioned, only one of them is chosen at a time in predicting a coding frame. The choice is signaled explicitly in the bitstream. Depending on the buffer size, more long-term key frames may be buffered. 

In passing, we stress that our current prediction structure is just one specific implementation. Our framework is able to accommodate other prediction structures. For example, how to mark and select long-term key frames is an encoder issue. The long-term key frame can even be the very first intra-frame. Likewise, the short-term reference frame can also be a decoded frame other than the previously decoded frame.


\subsection{Long- and Short-Term Temporal Prediction}
\label{sssec:Motion_Overhead}
The LSTP fuses the long- and short-term reference frames to generate the temporal predictor $x_c$ and condition signals $C^{1,2,3}$ for conditional residual coding (see Fig.~\ref{fig:OverView} (b)). The process involves (1) updating the temporally warped features $F^{1,2,4}_{key}$ of the long-term key frame as $\bar{F}^{1,2,4}_{key}$ based on the features $F^{1,2,4}_{pre}$ of the short-term reference frame and (2) fusing $\bar{F}^{1,2,4}_{key},F^{1,2,4}_{pre}$ via our MCNet, which has a GridNet-like structure~\cite{gridnet}. The former aims to complement the long-term key frame features with the short-term ones. The latter is to generate the temporal predictor $x_c$ and condition signals $C^{1,2,3}$ to perform conditional residual coding. The long-term key frame, which appears before the short-term reference frame in coding order, typically includes fewer temporal cascading errors and exhibits higher quality. However, it has a longer prediction distance compared to the short-term reference frame, suggesting that the long-term key frame may not form a good prediction of the coding frame in regions where motion estimates are unreliable. Motivated by these observations, we introduce a multi-scale gating mechanism to update $F^{1,2,4}_{key}$ with $F^{1,2,4}_{pre}$ in a spatially adaptive manner. The updating process is given by
\begin{equation}
\label{eq:adaptive_gate}
    \bar{F}^{i}_{key} = \gamma^{(i)} \odot F^{i}_{key} + (1-\gamma^{(i)}) \odot F^i_{pre},
\end{equation}
where $\gamma^{(i)} \in \mathbb{R}^{1 \times H/i \times W/i}, \; i \in \{ 1, 2, 4 \}$ represents the spatial-wise gating signal at scale $i$, and $\odot$ represents the element-wise multiplication. These spatial-wise masks $\gamma^{(i)}$ are generated by the spatial gate predictors. The elements of $\gamma^{(i)}$ are real values between 0 and 1, indicating the degree of reliability between the two reference sources.

\input{section/3-2_CodingStructure}


\subsubsection{Online Selection of Long-Term Key Frames}
\input{section/4-1_Pred_Struct}

When the long-term key frame is selected adaptively to encode $x_t$, we conduct an exhaustive search of the best long-term key frame by minimizing the per-frame rate-distortion cost:
\begin{equation}
    L^k = \lambda \cdot D(x_t, \hat{x}_t(\hat{x}_{t-1}, x^k_{key})) + R^k_t,
\end{equation}
where D measures the distortion between $x_t$ and its reconstruction $\hat{x}_t(\hat{x}_{t-1}, x^k_{key})$, which utilizes $\hat{x}_{t-1},x^k_{key}$ as the short-term and long-term reference frames, respectively, $R^k_t$ is the total bitrate, and $\lambda$ ranges between 228 and 1626, depending on the target bitrate. $k=1,2$ denotes the key frame index in the long-term section. The choice of the long-term key frame is signaled in the bitstream. 
\subsection{Alternative prediction structures}
We investigate several alternative prediction structures (see Fig.~\ref{fig:pred_struct}). These include (i) ``Short-Short ($SS$),'' as introduced previously, (ii) ``Two Previous ($TP$),'' which predicts a current frame from the last two previously decoded frames, and (iv) ``Long-Long ($LL$),'' which predicts from the last two long-term key frames. The ``Long-Short ($LS$)'' corresponds to our MH-LVC-1 prediction scheme, which has one short-term reference frame and one long-term key frame. Note that all the prediction structures share the same network weights trained solely for the $LS$ prediction.


%% file: section/3-1_OverView.tex
\begin{figure*}[htb]
    \centering
    \includegraphics[width=0.8\textwidth]{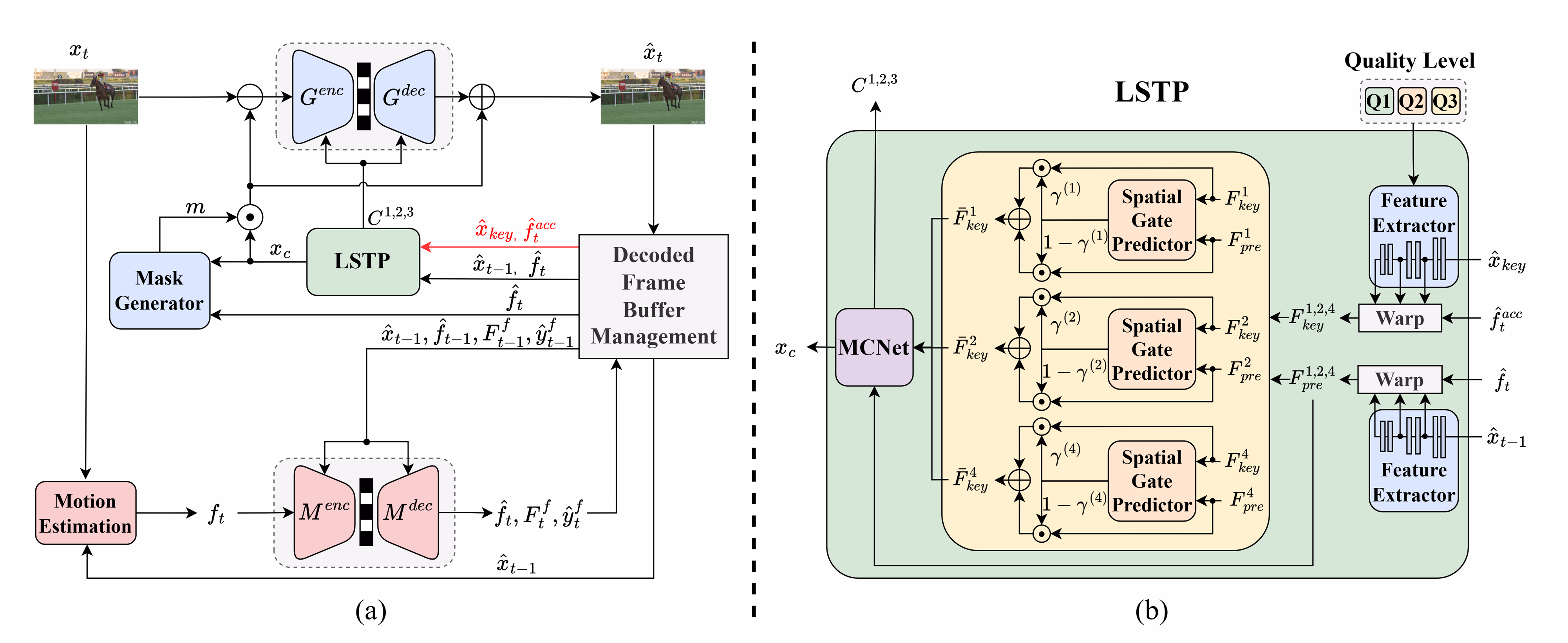}
    \vspace{-1.5 em}
    \caption{(a) The framework of the proposed MH-LVC. (b) The design details of the long- and short-term temporal prediction module (LSTP), which uses the quality level (Q1, Q2, Q3) to signal and adjust the decoded frame quality. $\gamma^{(i)},i=1,2,4$ are the gating signals used to update the long-term key frame features $F^{1,2,4}_{key}$.}
    \label{fig:OverView}
    \vspace{-2 em}
\end{figure*}

%% file: section/3-2_CodingStructure.tex
\begin{figure}[tbp]
    \centerline{\includegraphics[width=0.4\textwidth]{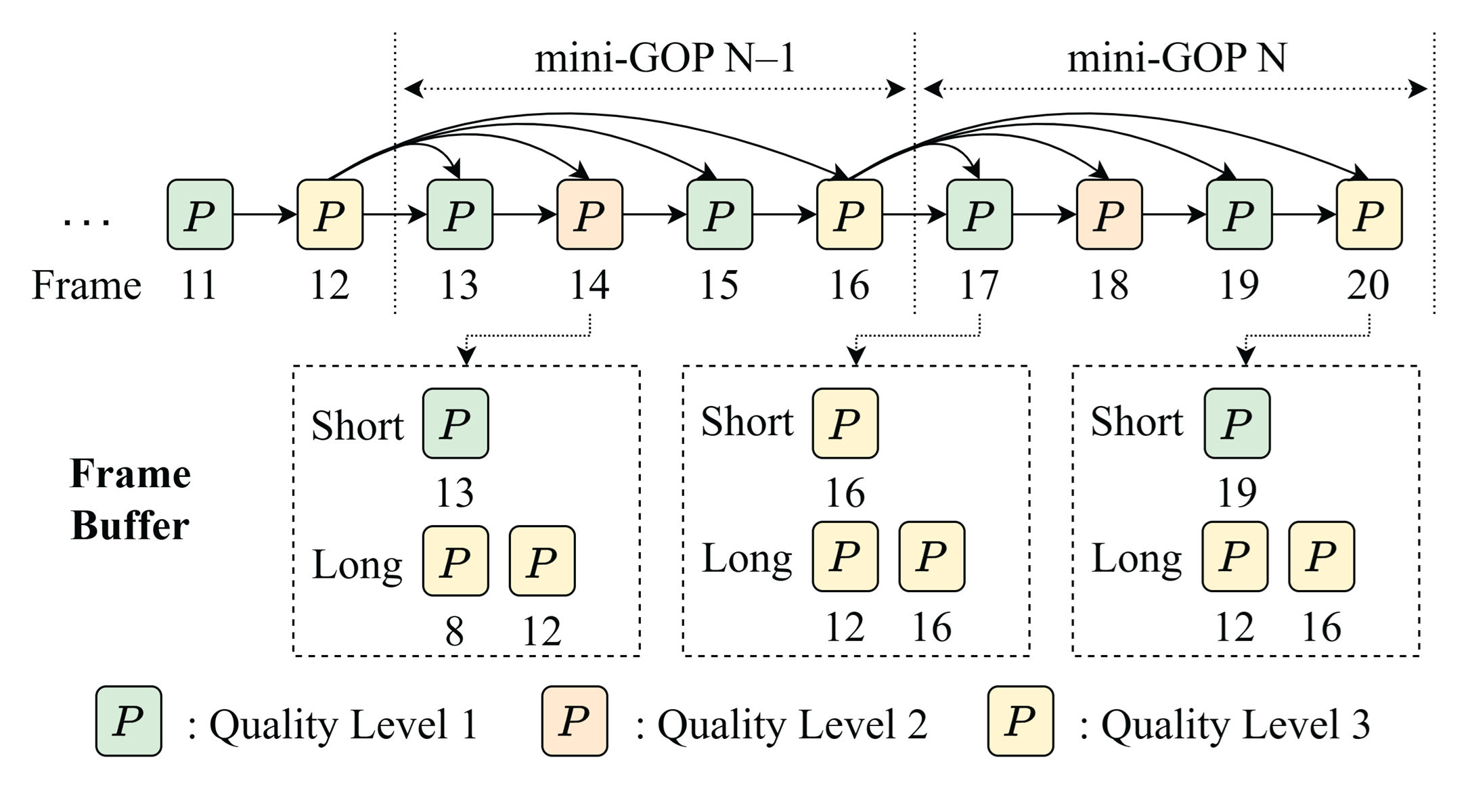}}
    \vspace{-0.5 em}
    \caption{Illustration of the coding order and temporal prediction structure, and how the contents of the decoded frame buffer evolve over time. The larger the quality level, the better the decoded image quality is intended to be. To avoid clutter, the long-term prediction is visualized only for the most recently decoded long-term key frame.}
    \label{fig:CondingStructure}
    \vspace{-1.em}
\end{figure}

%% file: section/4-1_Pred_Struct.tex
\begin{figure}[tbp]
    \centerline{\includegraphics[width=0.4\textwidth]{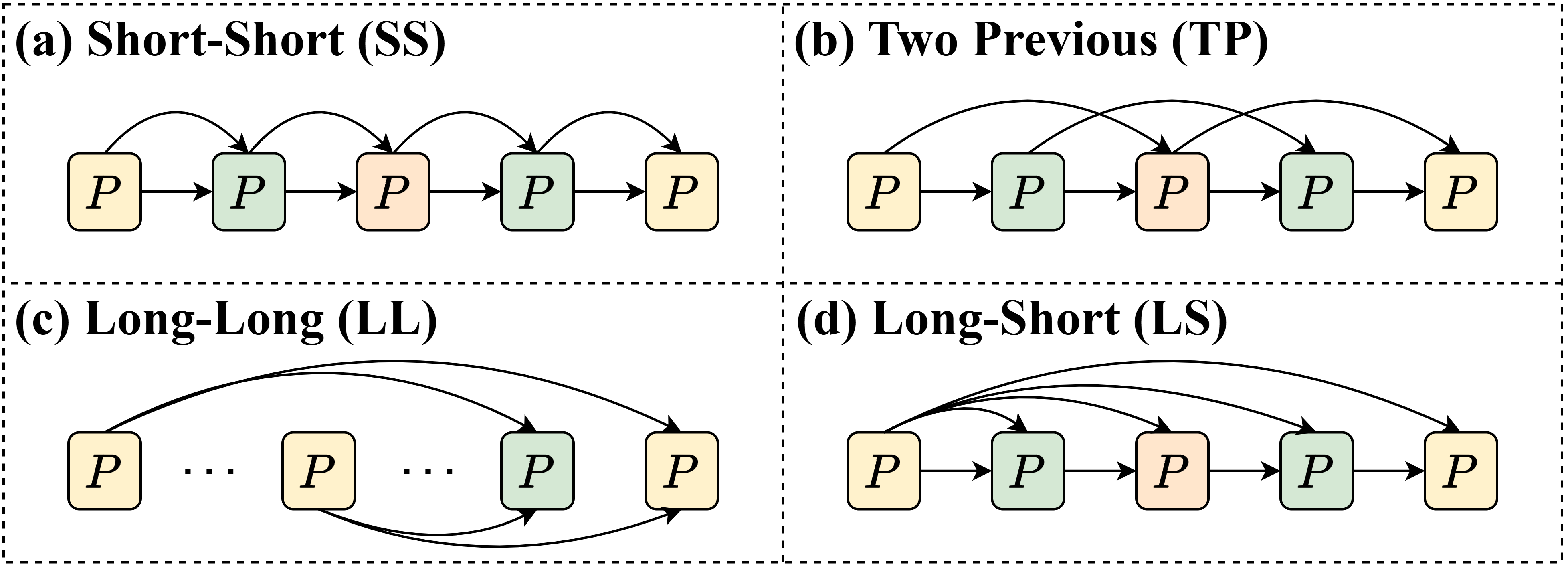}}
    \caption{Multi-hypothesis prediction structures. $LS$ corresponds to our MH-LVC-1.}
    \label{fig:pred_struct}
    \vspace{-1em}
\end{figure}

%% file: section/4_Result.tex
\section{Experimental Results}
\label{sec:results}

\label{sec:Evaluation_methodologies}

\noindent\textbf{Training details.} 
We train our model on the Vimeo-90k~\cite{vimeo} dataset, which includes 91,701 7-frame sequences of size $448 \times 256$. We select the first 5 frames in each sequence for training and randomly crop frames into $256 \times 256$ patches. We adopt the Adam~\cite{adam} optimizer and construct the training objective as follows: 
\begin{equation}
    \label{eqa:training_object}
    L = \frac{1}{4}\sum^{5}_{t=2}{\lambda \cdot w_t \cdot D(x_t, \hat{x}_t) + R_t},
\end{equation}
where $t,D(x_t,\hat{x}_t),R_t$ represent the frame index, distortion and bitrate, respectively. $\lambda$ is chosen to be 1626 in training a single-rate model. To impose a quality structure among coded frames for better bit allocation~\cite{dcvcdc,dcvcfm}, we choose $w_t$ as $\{ 1.2, 0.5, 1.2, 0.9 \}$ for $t=2$ to $t=5$ to form a nested quality structure. The larger the $w_t$, the more heavily the decoded frame quality is weighted. Due to limited compute resources, we do not use long training sequences, which were found to benefit some state-of-the-art learned codecs~\cite{dcvcfm}. More training details are provided in the supplementary document.

\noindent\textbf{Evaluation Methodologies.} 
Following the common test protocol~\cite{dcvcdc}, we evaluate our models on widely used datasets, including UVG~\cite{uvg}, HEVC Class B ~\cite{hevcctc}, and MCL-JCV~\cite{mcl}. For all the test sequences, we adopt BT.601~\cite{ffmpeg} to convert them from YUV420 to RGB444. We note that coding YUV420 content requires special care and is a separate task, which will be addressed in our future work. We set the intra period to 32, and encode the first 96 frames for each test sequence. For the MCL-JCV~\cite{mcl} dataset, we exclude screen content sequences from our evaluation and use intra coding at scene cuts. We thus crop all the test sequences to meet this requirement and apply the same setting to all the competing methods for a fair comparison. We report BD-rate savings~\cite{bdrate_old} in terms of peak signal-to-noise ratio (PSNR) in the RGB domain and bits-per-pixel (bpp). The negative and positive BD-rate numbers represent the rate reduction and inflation, respectively.


\subsection{Ablation Experiments}
\label{ssec:ablation}


\input{table/Ablations/MRF_per_seq}

\subsubsection{Long-term Key Frames}
For more insights into the rate-distortion benefits of long-term key frames, Table~\ref{tab:MRF_seq} presents the per-sequence BD-rate savings of our MH-LVC with one (MH-LVC-1) and two (MH-LVC-2) long-term key frames, respectively. In an effort to align the compression backbone and the other components to the extent possible, we use the $SS$ prediction structure, whose multi-hypotheses reference to the most recently decoded frame twice yet with the same optical flow map \textcolor{black}{(described in the supplementary)} as the anchor. This configuration simulates the effect of the single-hypothesis prediction with our two-hypothesis framework by referencing the most recently decoded frame twice yet with the same optical flow map. We chose it as the anchor rather than the existing single-hypothesis learned codecs such as DCVC~\cite{dcvc} and MaskCRT~\cite{maskcrt}, since their components are not exactly the same as those of our MH-LVC. Moreover, the three prediction structures share the same network weights. Thus, this anchor is preferable to the challenges of comparing fundamentally dissimilar entities. 

We observe that both MH-LVC-1 and MH-LVC-2 achieve considerable rate savings over $SS$. As expected, incorporating an additional long-term key frame enhances coding efficiency. MH-LVC favored sequences with simple motion (e.g.~BQTerrace), where the motion estimates between the coding frame and its long-term key frames are relatively more reliable. In complex-motion sequences (e.g.~ShakeNDry and Cactus), where the motion estimates are unreliable, MH-LVC shows promising coding gain. We remark that part of the gain of both MH-LVC-1 and MH-LVC-2 comes from the fact that the network weights are trained with the prediction structure combining the long- and short-term reference frames rather than $SS$. In general, combining multiple prediction hypotheses is an effective way to cope with motion uncertainty. From Table~\ref{tab:MRF_seq}, MH-LVC-2 performs better than MH-LVC-1. Having more reference choices allows the encoder to optimize the prediction structure based on the video content.

\subsubsection{Spatially Adaptive Feature Fusion}
\label{subsubsec:AMRFS}

\input{table/Ablations/W_WO_Gating}

Table~\ref{tab:Ada_gate} presents the BD-rate comparison to justify our spatially adaptive feature fusion (see Fig.~\ref{fig:OverView} (b)). As shown, with our spatial gating, significant bitrate savings are achieved across HEVC datasets. This suggests that some regions in the long-term key frame features $F^{1,2,4}_{key}$ should be refrained from prediction due to unreliable motion estimates. To further justify our design choice, we replace the gating mechanism in LSTP with simple convolutional blocks for feature fusion (w/ CNN Fusion in Table~\ref{tab:Ada_gate}). To ensure a fair comparison, we maintain computational complexity similar to that of Spatial Gate Predictor (see Fig.~\ref{fig:OverView} (b)). We observe that CNN-based fusion produces less favorable results, highlighting the importance of fusing feature maps by considering their reliability. Fig.~\ref{fig:mask_vis} further visualizes the highest-resolution mask $\gamma^{(1)}$ and prediction residues for $Jockey$. Two temporal prediction cases are compared. Recall that $\gamma^{(i)}$ ranges between 0 and 1, and indicates the spatial adaptive weighting for $F^{1,2,4}_{key}$ (see Fig.~\ref{fig:OverView} (b)). From Fig.~\ref{fig:mask_vis} (a), $\gamma^{(1)}$ weights lightly the key-frame features around the racing horse where dis-occlusion occurs. The unreliable motion estimates in the dis-occluded region are confirmed by the larger prediction residues. In the region with dis-occlusion, our gating mechanism learns to update $F^{1,2,4}_{key}$ by relying more heavily on features $F^{1,2,4}_{pre}$ from the short-term frame (see Fig.~\ref{fig:OverView} (b)). This stresses the importance of performing spatially adaptive fusion at the sub-frame level. From Fig.~\ref{fig:mask_vis} (b), when both prediction hypotheses come from the same short-term reference frame with the same flow map, $\gamma^{(1)}$ predicts values close to 0.5 across all the entire region.

\input{table/Ablations/3-3_Mask_Visualization}

\input{section/4-2_Per_Frame}

\input{table/Compare_SOTA/RD_Curve_I32}
\input{table/Compare_SOTA/RD_Curve_INF}

\subsubsection{Cascading Errors}
\textcolor{black}{As the long-term key frame at the highest quality level can preserve significantly more details, it helps mitigate temporal cascading errors. To validate this claim, we examine the PSNR profiles of $LS$ and $TP$, noted that $TP$ only references to the last two previously coded frames.} Fig.~\ref{fig:psnr_profiles} shows that our $LS$ is more effective than $TP$ in mitigating temporal cascading errors. The results stress the importance of incorporating both long- and short-term reference frames.

\subsection{Comparison with the State-of-the-Art Methods}
\label{subsec:SOTA}
We compare our MH-LVC with the state-of-the-art traditional and learned video codecs. We test their performance under an intra-period of 32 and of infinity (i.e. intra-period = -1). The traditional codecs include HM-16.25~\cite{hevc} and VTM-17.0~\cite{vvc} with \emph{encoder\_lowdelay\_main\_rext.cfg} and \emph{encoder\_lowdelay\_vtm.cfg}, respectively. Both configurations enable generalized B-frame coding, for a fair comparison with MH-CRT. Following~\cite{dcvcdc}, HM and VTM encode the input video in YUV444 format, with the decoded frame quality measured in the RGB domain. 

The learned codecs include B-CANF~\cite{bcanf}, DCVC-TCM~\cite{tcmli}, DCVC-HEM~\cite{hemli}, DCVC-FM~\cite{dcvcfm}, and MaskCRT~\cite{maskcrt}. Note that B-CANF is the state-of-the-art B-frame codec. It adopts a hierarchical B-frame prediction structure with intra-period 32, formulating the predictor based on both the past and future reference frames. MaskCRT is a P-frame codec that relies solely on the previously decoded frame for IPPP prediction. Although the DCVC series of works~\cite{tcmli, hemli, dcvcfm} are generally considered to be P-frame codecs, their temporal context mining aggregates latent information from all the past frames.

\subsubsection{Rate-distortion Comparison} 
\label{subsubsec:RD_Comparison}
\noindent \textbf{Intra-period 32.} Fig.~\ref{fig:RD_I32} presents the compression results for intra-period 32. We make the following observations. (1) With a finite intra-period 32, MH-LVC-2 performs similarly to MH-LVC-1 on most datasets. (2) Both MH-LVC variants perform close to DCVC-FM on most datasets, but with a much lower buffer size (as will be detailed in Section~\ref{subsubsec:complexity_comparison}). (3) Both MH-LVC variants surpass B-CANF, although adopting uni-directional temporal prediction. (4) They also outperform MaskCRT and DCVC-TCM considerably.

\noindent \textbf{Infinite Intra-period.} Fig.~\ref{fig:RD_INF} further presents results for an infinite intra-period. We follow~\cite{dcvcfm} to encode the first 96 frames for each test sequence. In this setting, (1) MH-LVC-2 significantly outperforms MH-LVC-1 on most datasets, suggesting that the additional long-term reference frame effectively mitigates cascading errors. Furthermore, (2) MH-LVC-2 performs comparably to DCVC-FM on UVG and HEVC-B, although it performs worse than DCVC-FM on MCL-JCV. It is important to note that both MH-LVC-1 and MH-LVC-2 are trained on 5-frame GOPs to reduce training time, whereas DCVC-FM is trained on 32-frame GOPs. Training on large GOPs is expected to benefit the coding performance when the intra-period is infinite. However, MH-LVC-2 still manages to achieve good coding performance. The results suggest a new research direction that has not yet been explored for learned video codecs.

\subsubsection{Complexity Analysis} 
\label{subsubsec:complexity_comparison}

Table~\ref{tab:sota_complexity} summarizes the complexity aspects of several learned codecs with uni-directional temporal prediction. Our MH-LVC-2 (and likewise MH-LVC-1) buffers much fewer full-resolution feature maps than the DCVC family. For instance, DCVC-FM requires 48 full-resolution (i.e.~the same spatial resolution as the input image) feature maps, 1 reconstructed frame, and the latent prior, requiring the equivalent of 51.75 full-resolution feature maps. All these feature maps must be retrieved in decoding a video frame. In contrast, MH-LVC-2 stores 3 reconstructed frames, 2 reconstructed flow maps, auxiliary information for the motion codec, requiring the equivalent of 19.25 full-resolution feature maps. More importantly, it restricts the number of reference frames to be used for temporal prediction at a time to two. In decoding a video frame, MH-LVC-2 retrieves the equivalent of 19.25 full-resolution feature maps, as compared to 14.25 with MH-LVC-1.

In terms of the number of operations, MH-LVC-2 and MH-LVC-1 share the same decoding kMAC/pixel, which falls between those of DCVC-FM and DCVC-HEM. It is important to note that DCVC-FM has put significant effort into reducing the model's size and encoding/decoding kMAC/pixel. The other methods have not undergone similar levels of optimization and are expected to benefit similarly from their network optimization techniques. As anticipated, MH-LVC-2 has a much higher encoding kMAC/pixel than MH-LVC-1 (and the competing methods) due to the exhaustive search of the long-term key frame. This encoder issue can be addressed by a fast frame-selection algorithm, which is among our future work. High encoding complexity is also common in traditional codecs (e.g.~VTM and HM), particularly when there are multiple encoding options to choose from. In terms of decoding runtimes, MH-LVC-1 and  MH-LVC-2 have similar decoding times (0.54-0.56s) to DCVC-FM (0.47s). Additionally, MH-LVC-1 shows only a slightly higher encoding time (0.73s) compared to DCVC-FM (0.6s). These results imply a high degree of parallelism inherent in the encoding/decoding processes of MH-LVC.

To conclude, Table~\ref{tab:complexity_DCVCFM} summarizes the rate-distortion-complexity trade-offs of our MH-LVC and DCVC-FM. This work explores uncharted territory for learned video codecs by incorporating long- and short-term reference frames to improve coding efficiency at the cost of a much lower decoded frame buffer size. In terms of decoding complexity, both MH-LVC-1 and MH-LVC-2 exhibit similar decoding runtimes to DCVC-FM, although their decoding kMAC/pixel is slightly higher. In terms of encoding complexity, MH-LVC-2 incurs a much higher encoding time and kMAC/pixel due to an exhaustive search of the best long-term key frame. This encoder issue can be further addressed by a fast frame-selection algorithm in the future. We note that reducing kMAC/pixel and runtime is important and can be achieved through various network optimization and implementation techniques (without significantly changing the algorithm itself). However, the size of the decoded frame buffer is more intrinsic to the algorithm.

\input{table/Compare_SOTA/SOTA_Complexity}

\input{table/Ablations/Compare_DCVCFM}

\vspace{-0.5 em}

\subsection{Comparison with Implicit Buffering Strategies}
\label{subsubsec:implicit_buffering_comparison}

Our MH-LVC adopts an explicit buffering strategy by storeing explicitly decoded frames for temporal prediction. In contrast, recent DCVC codecs, e.g. DCVC-FM and DCVC-TCM, implement an implicit buffering strategy, relying entirely on learning latent features for temporal prediction. Table~\ref{tab:compare_TCM} presents the BD-rate comparison between the explicit and implicit buffering strategies with an intra\_period of 32. For a fair comparison, we implement the implicit buffering strategy of DCVC-TCM~\cite{tcmli} within the same conditional residual video coding framework and follow the same training strategy as MH-LVC-1. Its detailed network architecture is provided in the supplementary material.

From Table~\ref{tab:compare_TCM}, our MH-LVC-1, which incorporates two reference frames in the frame buffer for temporal prediction, outperforms TCM, which learns and stores only latent features for the same purpose. In this experiment, both exhibit comparable encoding and decoding kMAC/pixel because they share most of the coding components. Also, we follow~\cite{dcvctcm} to allocate a large buffer for TCM, which has a size much larger than that (14.25) of MH-LVC-1.

\input{table/Ablations/Compare_TCM}

%% file: table/Ablations/MRF_per_seq.tex
\begin{table}[tbp]
\centering
\caption{The per-sequence BD-rates on UVG and HEVC-B datasets: MH-LVC-1 versus MH-LVC-2. The anchor is $SS$, which simulates the effect of the single-hypothesis prediction with our MH-LVC by referencing the most recently decoded frame twice yet with the same optical flow map.}
\vspace{-0.5 em}
\footnotesize
\begin{tabular}{cccc}
\toprule

\multirow{2}{*}{Dataset} & \multirow{2}{*}{Sequence} & \multicolumn{2}{c}{BD-rate (\%) PSNR-RGB} \\

\cmidrule(r){3-4}
& & MH-LVC-1 & MH-LVC-2 \\
\midrule
\multirow{7}{*}{UVG} 
& Beauty & -8.5 & -12.9 \\
& Bosphorus & -7.9 & -9.7 \\
& HoneyBee & -8.4 & -11.6 \\
& Jockey & -8.2 & -9.6 \\
& ReadySteadyGo & -8.7 & -10.4 \\
& ShakeNDry & -14.5 & -18.8 \\
& YachtRide & -11.3 & -14.6 \\
\cmidrule{2-4}
& \textbf{Average} & \textbf{-9.6} & \textbf{-12.5} \\
\midrule
\multirow{5}{*}{HEVC-B} 
& BasketballDrive & -11.5 & -13.9 \\
& BQTerrace & -14.2 & -18.3 \\
& Cactus & -17.7 & -20.9 \\
& Kimono1 & -10.5 & -13.6 \\
& ParkScene & -10.2 & -12.0 \\
\cmidrule{2-4}
& \textbf{Average} & \textbf{-12.8} & \textbf{-15.7} \\
\bottomrule
\label{table:MRF_seq}
\end{tabular}
\vspace{-3 em}

\label{tab:MRF_seq}
\end{table}

%% file: table/Ablations/W_WO_Gating.tex
\begin{table}[tbp]
\centering
\renewcommand{\arraystretch}{0.6}
\vspace{-1 em}
\caption{Ablation study of different feature fusion mechanisms in LSTP on the HEVC datasets. The anchor for BD-rate evaluation is the model without spatial gating (w/o Spatial Gating).}

\footnotesize
\begin{tabular}{c cccc}
\toprule
  & Class B & Class C & Class D & Class E     \\
\midrule
\multicolumn{1}{c}{w/o \; Spatial Gating}  & 0.0 & 0.0 & 0.0 & 0.0 \\
\midrule
\multicolumn{1}{c}{w/ \;\;\; Spatial Gating} & -4.2 & -5.9 & -5.5 &  -13.1  \\
\midrule
\multicolumn{1}{c}{w/  CNN-based Fusion} & -2.1 & -3.0 & -2.8 &  -9.1  \\
\bottomrule
\end{tabular}

\label{tab:Ada_gate}
\vspace{-2 em}
\end{table}

%% file: table/Ablations/3-3_Mask_Visualization.tex
\begin{figure}[t!]
\centering
\footnotesize
{
    \includegraphics[width=0.8\linewidth]{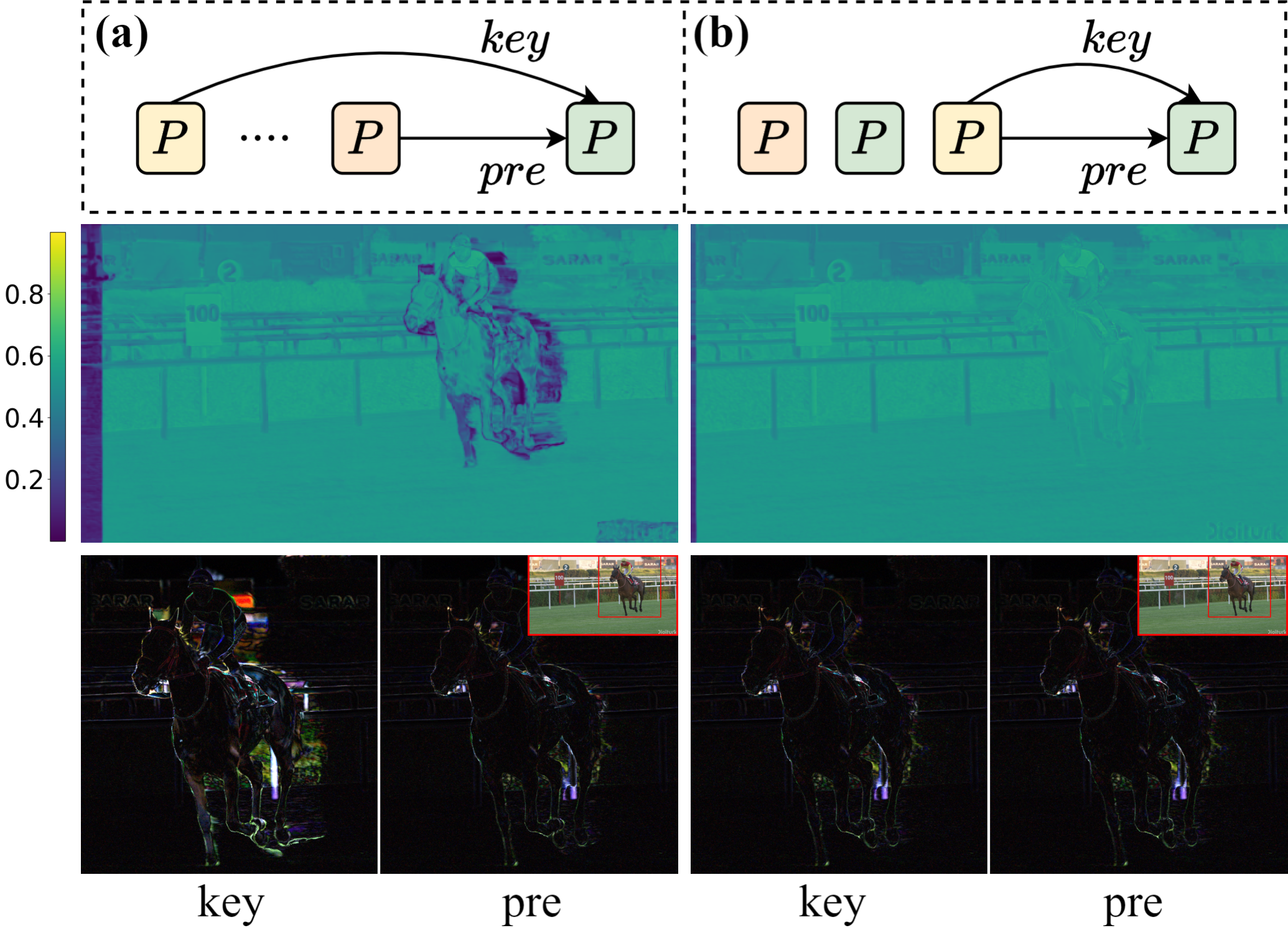} 
    \\
    
}

\vspace{-0.8 em}
\caption{Visualization of the gating signal $\gamma^{(1)}$ for two temporal prediction structures. (a) adopts both long- and short-term reference frames, and (b) has two predictors derived from the same short-term reference frame with the same optical flow map. The bottom row displays the prediction residues between the coding frame $x_t$ and its two motion-compensated reference frames $\hat{x}_{key}$ (denoted as key) and  $\hat{x}_{t-1}$ (denoted as pre).}

\vspace{-1 em}
\label{fig:mask_vis}

\end{figure}

%% file: section/4-2_Per_Frame.tex
\begin{figure}[t]
    \centering
        \includegraphics[width=0.4\textwidth]{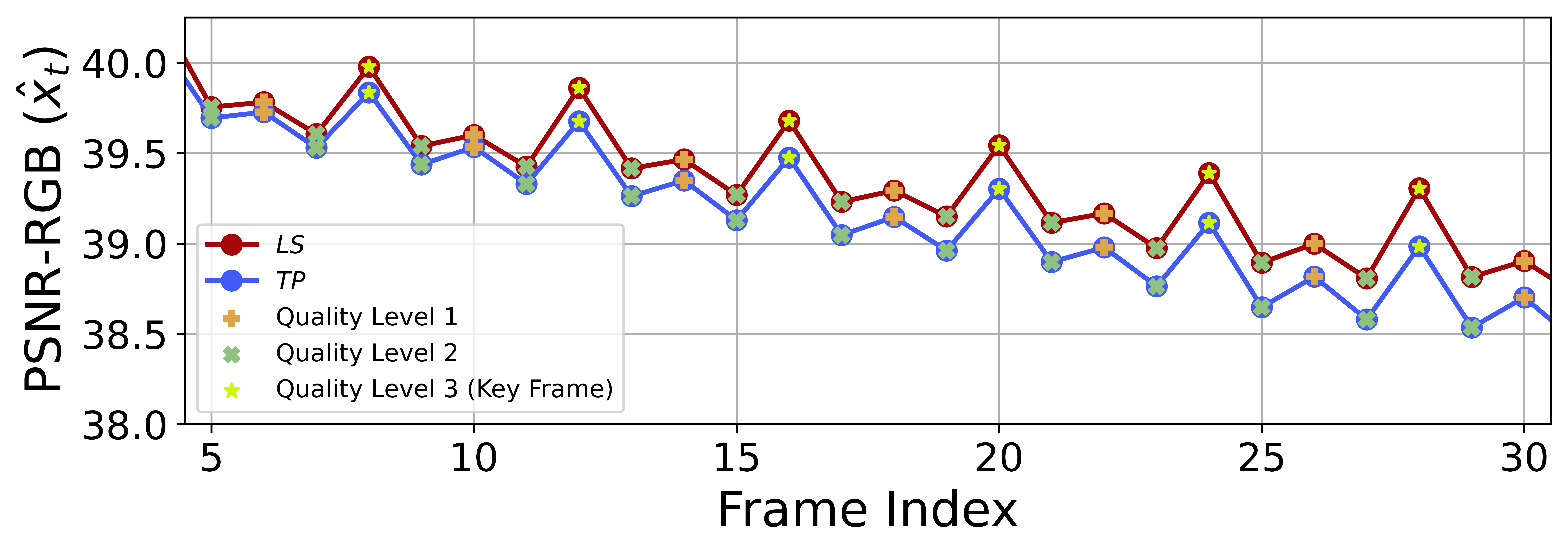}
    \vspace{-1. em}
    \caption{The per-frame PSNR for KristenAndSara: $LS$ versus $TP$. Their average bitrates are comparable.}
    \label{fig:psnr_profiles}
    \vspace{-1em}
\end{figure}

%% file: table/Compare_SOTA/RD_Curve_I32.tex
\begin{figure*}[tbp]
    \begin{center}
    \begin{subfigure}{0.31\linewidth}
        \centering
        \includegraphics[width=\linewidth]{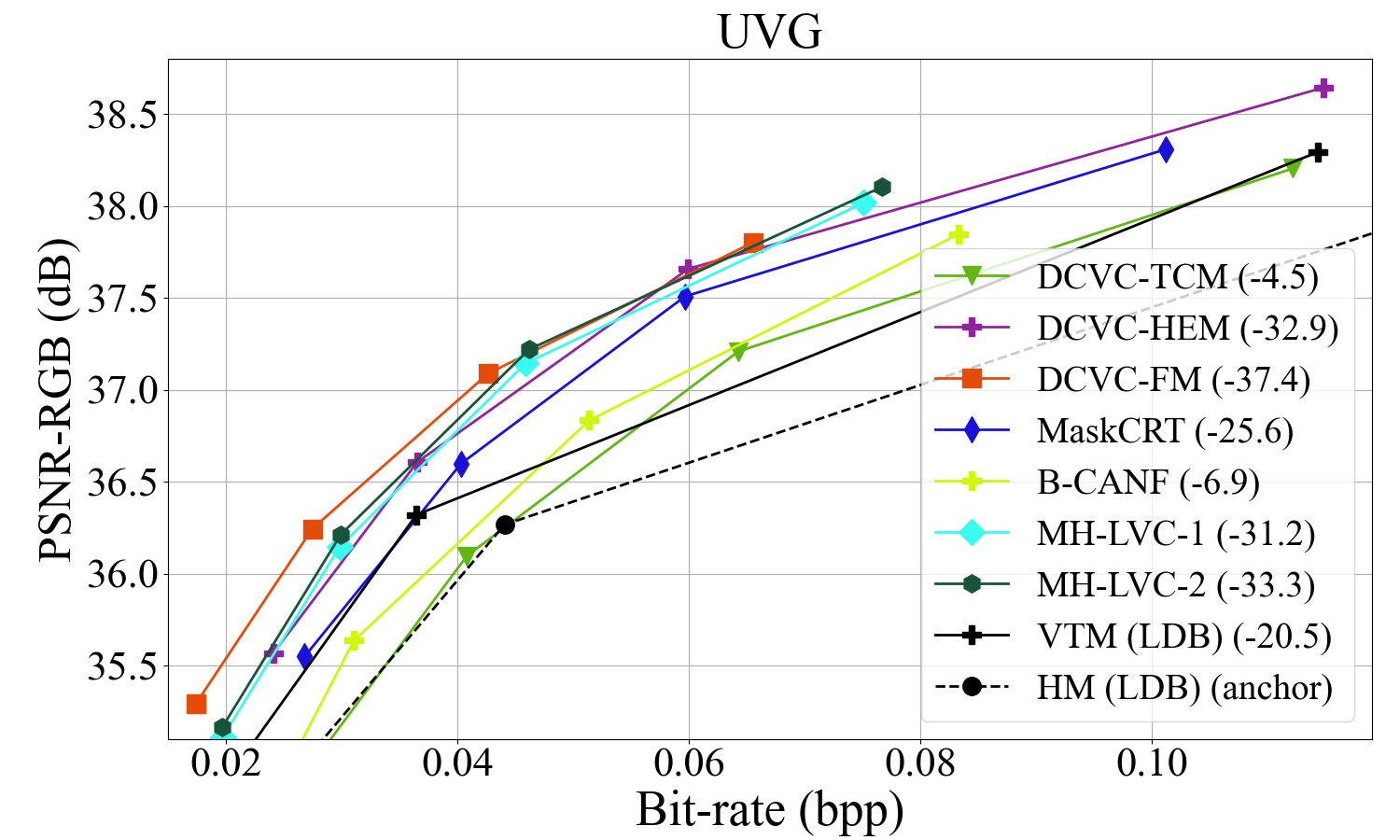}
    \end{subfigure}
    \begin{subfigure}{0.31\linewidth}
        \centering
        \includegraphics[width=\linewidth]{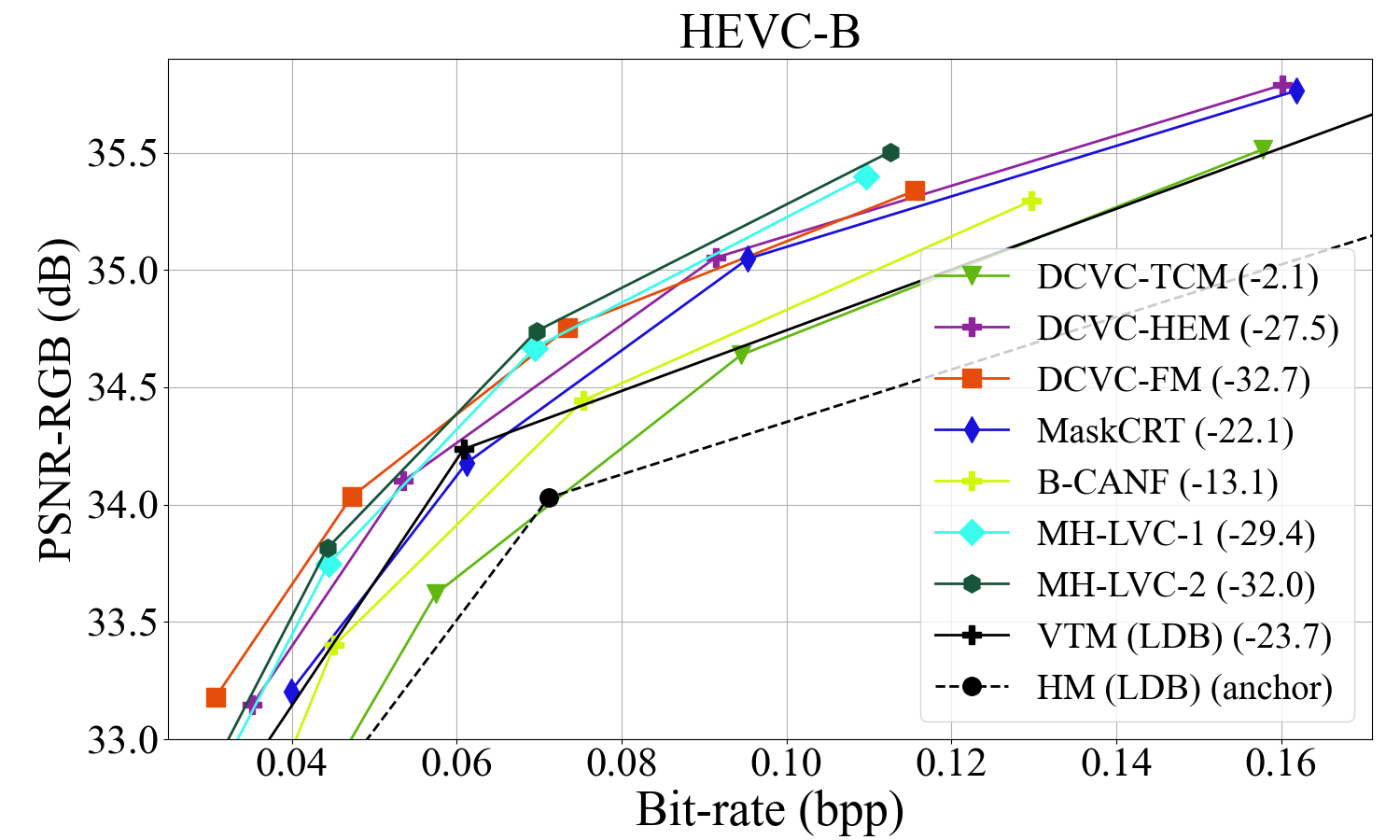}
    \end{subfigure}
    \begin{subfigure}{0.31\linewidth}
        \centering
        \includegraphics[width=\linewidth]{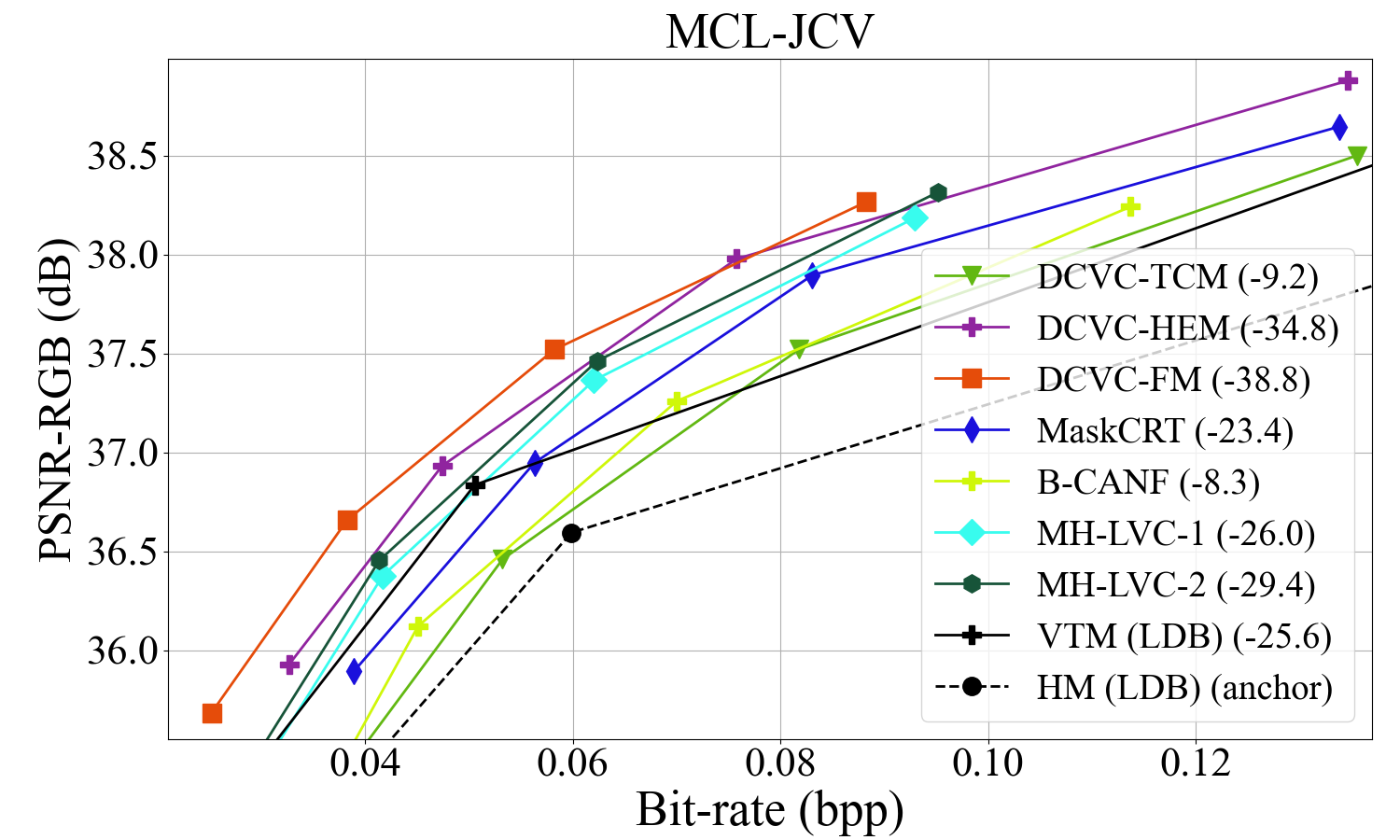}
    \end{subfigure}
    \vspace{-0.5 em}
    \caption{Rate-distortion performance comparison with intra-period 32. The numbers within the parentheses are BD-rates, with HM-16.25 (Low-delay B) serving as the anchor.}
    \label{fig:RD_I32}
    \end{center}
    \vspace{-1.5 em}
\end{figure*}

%% file: table/Compare_SOTA/RD_Curve_INF.tex
\begin{figure*}[t!]
    \begin{center}
    \begin{subfigure}{0.31\linewidth}
        \centering
        \includegraphics[width=\linewidth]{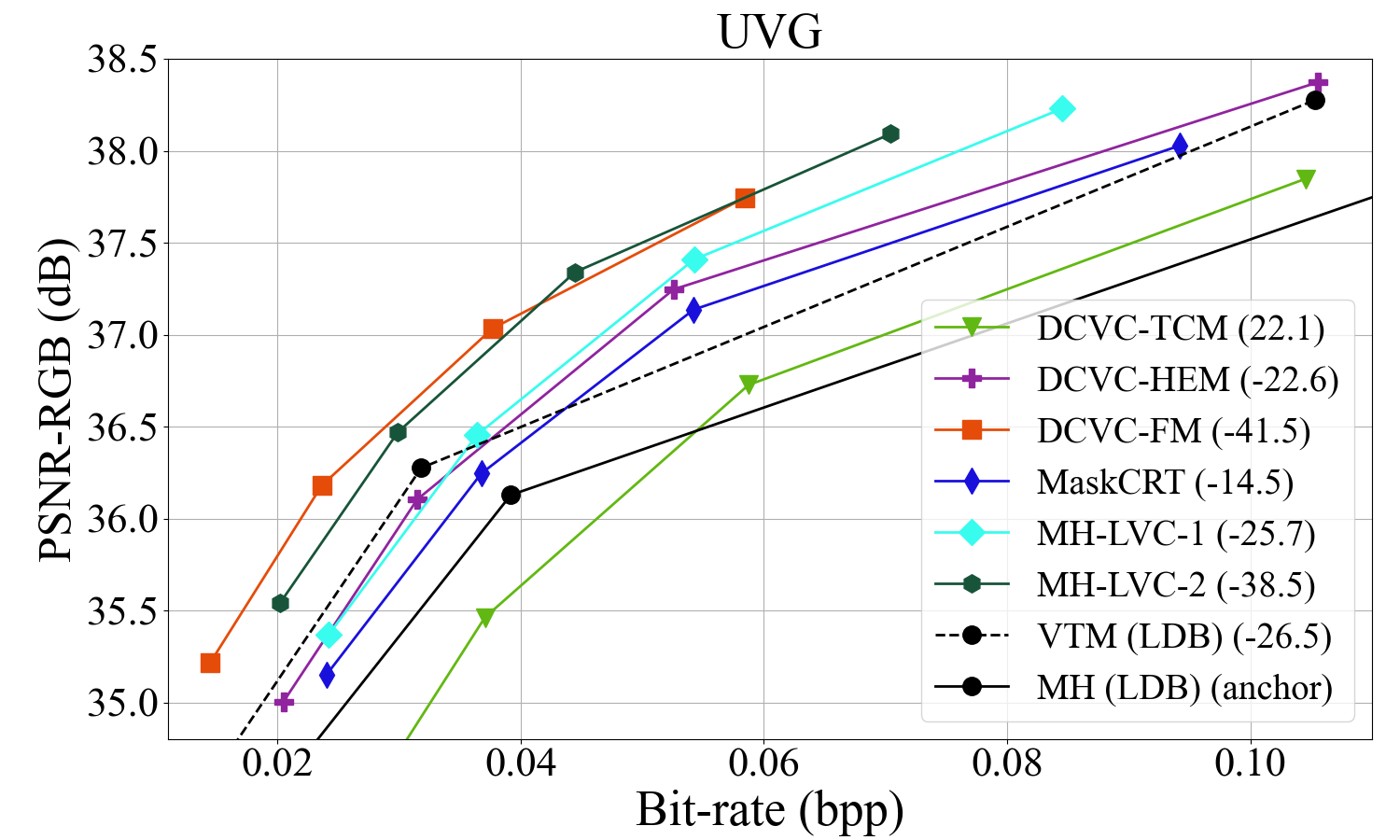}
    \end{subfigure}
    \begin{subfigure}{0.31\linewidth}
        \centering
        \includegraphics[width=\linewidth]{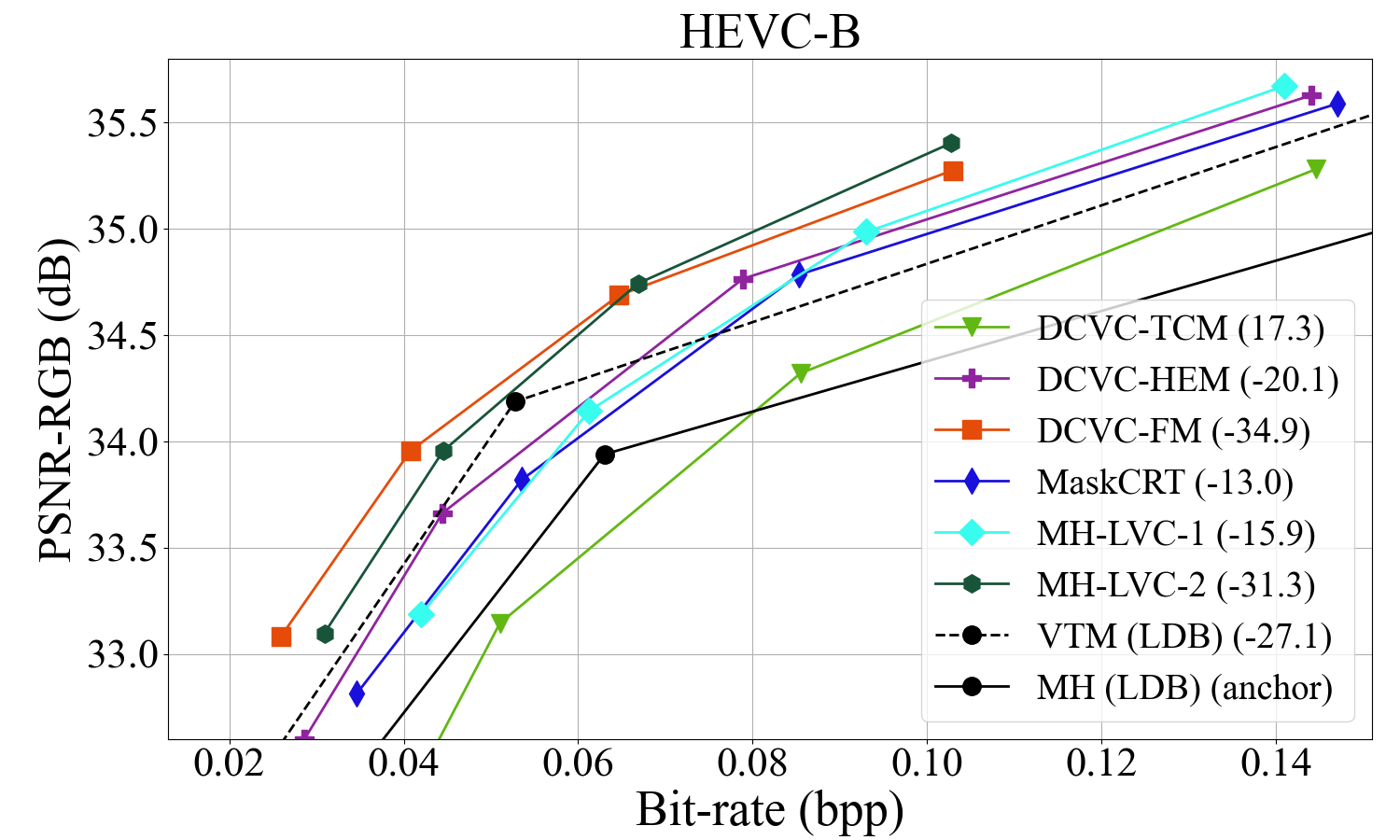}
    \end{subfigure}
    \begin{subfigure}{0.31\linewidth}
        \centering
        \includegraphics[width=\linewidth]{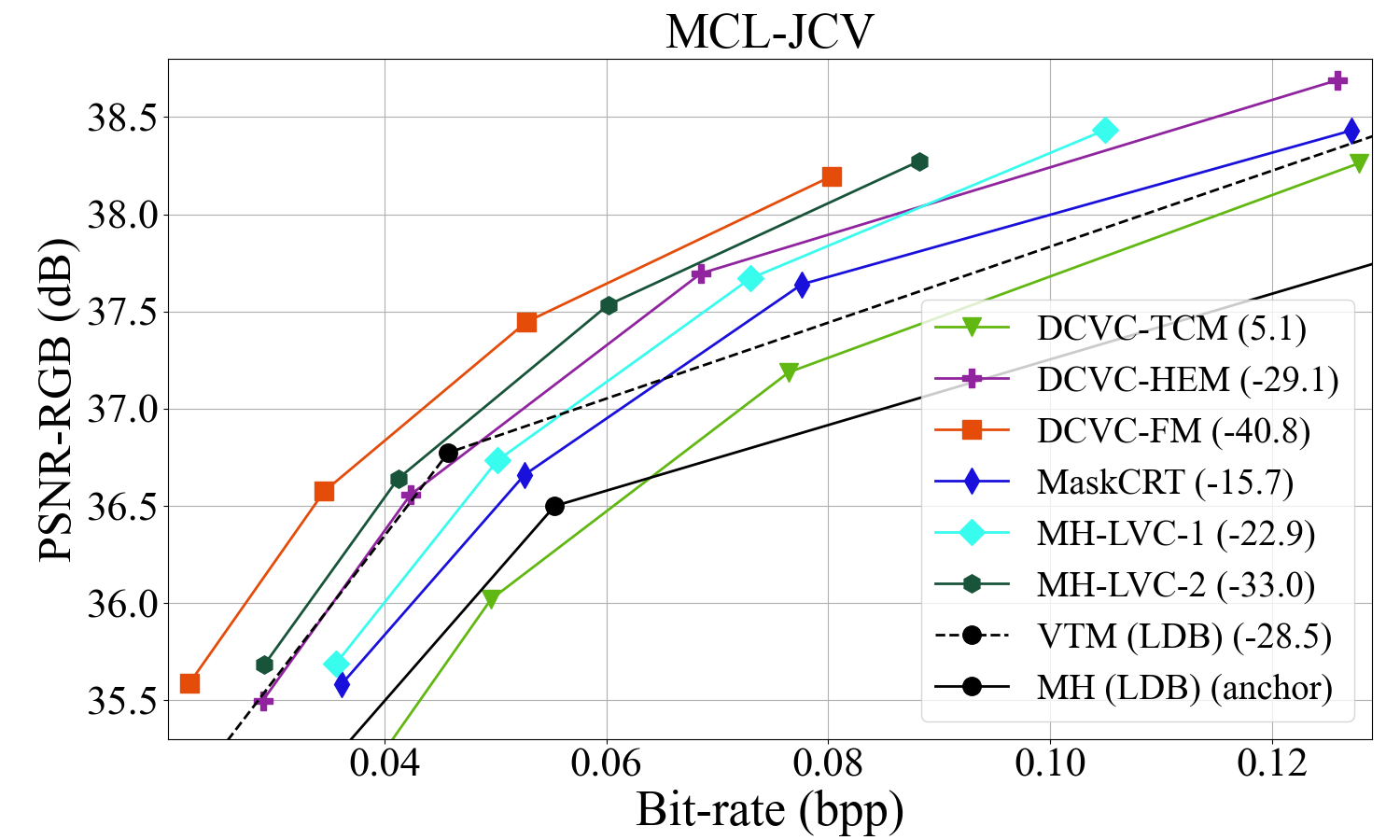}
    \end{subfigure}
    \caption{Rate-distortion performance comparison under an infinite intra-period. The numbers within the parentheses are BD-rates, with HM-16.25 (Low-delay B) serving as the anchor.}
    \label{fig:RD_INF}
    \end{center}
    \vspace{-3 em}
\end{figure*}

%% file: table/Compare_SOTA/SOTA_Complexity.tex
\begin{table}[t!]
\centering
\caption{Complexity comparison in terms of the model size, kMAC/pixel for encoding/decoding, and decoded frame buffer. The buffer size is measured in the number of full-resolution feature maps required to decode an input frame. Encoding/Decoding time is the average on HEVC-B.}
\footnotesize
\begin{tabular}{cccccc}
\toprule
Methods  & 
\begin{tabular}[c]{@{}c@{}}Model\\ Size (M)\end{tabular} &
\begin{tabular}[c]{@{}c@{}}Enc / Dec\\ kMAC/pixel\end{tabular} & 
\begin{tabular}[c]{@{}c@{}}Buffer \\ Size\end{tabular} &
\begin{tabular}[c]{@{}c@{}}Enc / Dec\\ time (s)\end{tabular}  \\ 
\midrule
 DCVC-TCM   & 10.7 & 1406\; /\;\; 917  & 67  & \textcolor{black}{0.61 / 0.38 }    \\ 
 DCVC-HEM   & 17.5 & 1662\; /    1243  & 67.63  & \textcolor{black}{0.57 / 0.34}\\ 
DCVC-FM    & 17.1 & 1137\; /\;\; 871  & 51.75 &  0.60 / 0.47  \\ 
MaskCRT    & 27.7 & 1401\; /\;\; 763  & 13   &  \textcolor{black}{1.83 / 1.04 }  \\ 
MH-LVC-1   & 16.0 & 1507 \; /    1088  & 14.25  &  0.73 / 0.54  \\   
MH-LVC-2   & 16.0 & 2611\; /    1088  & 19.25  & 1.27 / 0.56  \\
\bottomrule
\end{tabular}

\label{tab:sota_complexity}
\end{table}

%% file: table/Ablations/Compare_DCVCFM.tex
\begin{table}[tbp]

\centering
\footnotesize
\renewcommand{\arraystretch}{1.2}
\caption{The rate-distortion-complexity trade-offs of MH-LVC and DCVC-FM. }
\vspace{-1 em}
\begin{tabular}{l|c}
\hline

\multicolumn{1}{c|}{Evaluation} & Comparison                                          \\ \hline
Enc. kMAC/pixel             & MH-LVC-2 $\gg$ MH-LVC-1 $\textgreater$ DCVC-FM \\
Dec. kMAC/pixel             & MH-LVC-2 $\approx$ MH-LVC-1 $\gtrsim$ DCVC-FM            \\
Encoding time                   & MH-LVC-2 $\gg$ MH-LVC-1 $\gtrsim$ DCVC-FM \\
Decoding time                   & MH-LVC-2 $\approx$ MH-LVC-1 $\approx$ DCVC-FM                       \\
Buffer size                     & MH-LVC-1 $<$ MH-LVC-2 $\ll$ DCVC-FM   \\
Model size                     & MH-LVC-1 $=$ MH-LVC-2 $<$ DCVC-FM  
\\ \hline
BD-rate (Intra-period 32)                   & MH-LVC-1 $<$ MH-LVC-2 $\approx$ DCVC-FM                       \\
BD-rate (Intra-period -1)                   & MH-LVC-1 $\ll$ MH-LVC-2 $\approx$ DCVC-FM                       \\
\hline

\end{tabular}
\label{tab:complexity_DCVCFM}
\vspace{-2em}
\end{table}

%% file: table/Ablations/Compare_TCM.tex
\begin{table}[tbp]
\centering
\caption{Comparison of BD-rates between MH-LVC-1 and TCM~\cite{dcvctcm} within the same conditional residual coding framework. The intra-period is 32.}
\vspace{-1 em}
\setlength{\tabcolsep}{2.5 pt}
 \renewcommand{\arraystretch}{0.8}
\footnotesize
\begin{tabular}{c cccc c c}
\toprule
  & UVG & HEVC-B & MCL-JCV & \begin{tabular}[c]{@{}c@{}}Buffer\\ Size\end{tabular} &  \begin{tabular}[c]{@{}c@{}}Enc / Dec\\ kMAC/pixel\end{tabular}   \\
\midrule
\multicolumn{1}{c}{ \; MH-LVC-1} & 0.0 & 0.0  & 0.0 & 14.25 & 1507 / 1088  \\ 
\midrule
\multicolumn{1}{c}{ \; TCM} & 7.8 & 3.2 &  8.9 & 55 & 
1441 / 1021 
\\
\bottomrule
\end{tabular}

\label{tab:compare_TCM}
\vspace{-2 em}
\end{table}

%% file: section/5_Conclusion.tex
\vspace{-0.5em}
\section{Concluding Remarks}
\label{sec:conclusion}
\vspace{-0.5em}
This work presents an unexplored attempt at introducing an explicit buffering scheme to learned video coding. It allows both long- and short-term reference frames to be combined for multi-hypothesis temporal prediction in a conditional residual coding framework. Compared to the implicit buffering technique used in the recent conditional coding framework, our scheme achieves good coding performance at the cost of much smaller decoded frame buffer size. We expect that its rate-distortion performance can be further improved by using long training sequences, as is done in most competing methods. Currently, its encoding kMAC/pixel increases considerably due to an exhaustive search of the best long-term key frame. Developing a fast frame selection algorithm and performing structure optimization to reduce the decoding kMAC/pixel are among our future work.

\vspace{-1 em}

%% file: supp.tex






\clearpage
\beginsupplement




\maketitlesupplementary

\thispagestyle{empty}
\footnotetext[1]{Equal contribution.}
\input{section/A_Abstract}

\input{section/D_MiniGOP}
\input{section/F_Implementation_Details}

\input{section/G_Coding_Structure}
\input{section/E_TCM_overview}
\input{section/C_Alternative_pred}
\input{section/I_MRF}
\input{section/H_Traditional_Codecs}

\input{section/J_Compare_SOTA}

\input{section/K_More_Visualization}




%% file: section/A_Abstract.tex
This supplementary document provides the following additional materials and results to assist with the understanding of our MH-LVC:
\begin{itemize}
    \setlength{\itemsep}{0pt}
    \item Implementation details in 
    Section~\ref{sec:imple};
    \item Cascading and prediction errors in 
    Section~\ref{sec:first};
    \item Overview of Implicit Buffering Strategies in section~\ref{sec:implicit}
    \item Additional alternative Temporal Prediction Structures in section~\ref{sec:alter_pred}
    \item The number of long-term key frames in Section~\ref{sec:mrf};
    \item Command lines for VTM and HM in 
    Section~\ref{sec:traditional};
    \item Comparison with the state-of-the-art
methods in terms of MS-SSIM-RGB in 
    Section~\ref{sec:com_SOTA};
    \item More visualizations in 
    Section~\ref{sec:more_vis};
\end{itemize}

%% file: section/D_MiniGOP.tex
\section{Mini-GOP}
\label{sec:miniGOP}
Table~\ref{tab:supp_Mini_GOP_Period} examines the impact of the mini-GOP size on coding performance. These results justify our choice of mini-GOP 4.

%% file: section/F_Implementation_Details.tex
\section{Implementation Details}
\label{sec:imple}

\subsection{The Prediction Structure for Training}
\label{subsec:training}

\input{section/D-2_TrainTest_Structure}
We adopt a 5-frame training strategy due to limited compute resources. We remark that our scheme can benefit from training on large GOPs and long sequences.

Fig.~\ref{fig:supp_CodingStructure} (a) illustrates the temporal prediction structure during training. To create a quality structure among the decoded video frames, the weights $w_t$ of $\{1.2, 0.5, 1.2, 0.9\}$ are assigned as follows: (1) 1.2 for Frame 2 ($P^*$ frame), (2) 0.5 for Frame 3 with the quality level 2, (3) 1.2 for Frame 4 with the quality level 3, and (4) 0.9 for Frame 5 with the quality level 1. Notably, following a strategy similar to that of DCVC-DC~\cite{dcvcdc}, a specific feature extractor is trained to accommodate each of these weights. That is, in Fig.~\ref{fig:OverView} (b) of the main paper, the feature extractor of the long-term key frame $\hat{x}_{key}$ changes with the quality level. The $P^*$ frame right after the I-frame is unique in that it typically exhibits a much higher bitrate (and thus better decoded quality) than those of the remaining P-frames due to error propagation aware training~\cite{EPA}. In the setting with an infinite intra period, we enable $P^*$ frames periodically to mitigate temporal cascading errors. 

\input{table/Supp_Ablations/MiniGOP/Different_miniGOP}

To be as consistent with the prediction scenario at inference time as possible, we always use the most recently decoded frame as the short-term reference frame during training. Likewise, the I-frame, due to its higher quality, is always used as the long-term key frame. However, at inference time, the long-term key frame can be an I-frame, a $P^*$ frame, or a P-frame with the quality level 3.

\input{table/Supp_Training_Details/Training_Procedures}

\subsection{The Prediction Structures for Inference}
Fig.~\ref{fig:supp_CodingStructure} (b) illustrates our prediction structure at inference time and how the decoded frame buffer evolves over time under an intra period of 32. For forming a 4-frame mini-GOP, we follow a quality pattern similar to the hierarchical P prediction in traditional codecs. For the first mini-GOP where no prior key frames are available, the I-frame and $P^*$ frame are stored in the long-term section to serve as the long-term reference frames for the subsequent frames.

Fig.~\ref{fig:supp_CodingStructure} (c) illustrates the case with an infinite intra period, where we enable $P^*$ frames periodically to mitigate temporal cascading errors.

\subsection{Training Procedures}
Table~\ref{tab:supp_training_procedure} summarizes our training procedure. It begins with training the single-rate model, followed by fine-tuning it to arrive at the variable-rate model. The code will be made available for reproducibility upon the acceptance of the paper.

%% file: section/D-2_TrainTest_Structure.tex
\begin{figure*}[t]
    \centerline{\includegraphics[width=1\textwidth]{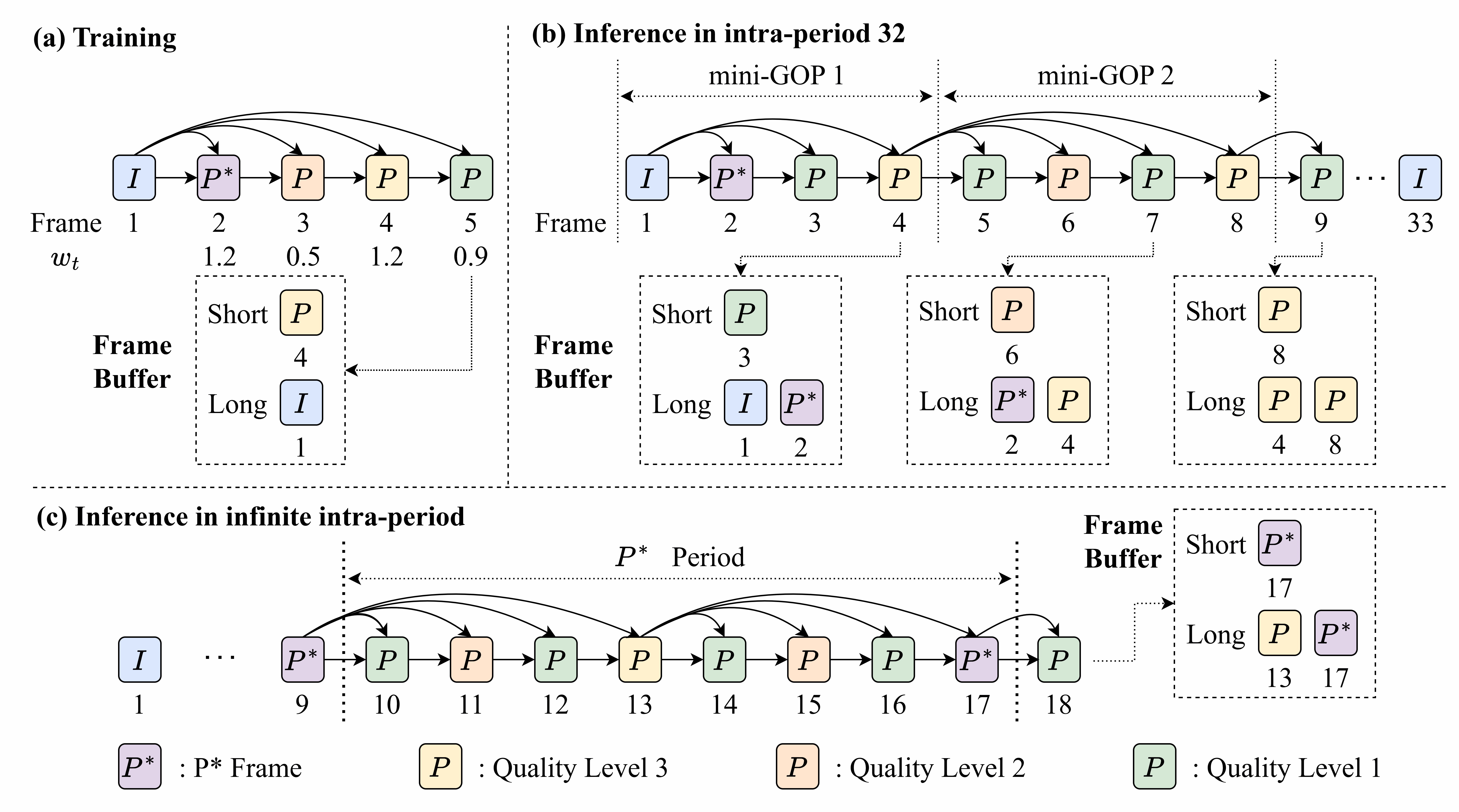}}
    \vspace{-1.0em}
    \caption{Illustration of the coding structures for (a) training with intra-period 32, (b) inference with intra-period 32, and (c) inference with an infinite intra-period.}
    \label{fig:supp_CodingStructure}
\end{figure*}

%% file: table/Supp_Ablations/MiniGOP/Different_miniGOP.tex
\begin{table}[tbp]
\centering
\setlength{\tabcolsep}{4.0pt}
\footnotesize
\caption{Ablation of different mini-GOP sizes. The anchor is $LS$ with a mini-GOP size of 4.}
\begin{tabular}{c cc}
\toprule
 \multicolumn{1}{c}{mini-GOP Period}  & HEVC-B & UVG \\
\midrule
\multicolumn{1}{c}{mini-GOP \;\;4} &  0.0 & 0.0 \\
\multicolumn{1}{c}{mini-GOP \;\;8} &  1.0 & 0.5  \\
\multicolumn{1}{c}{mini-GOP 12} &  2.5 & 1.3  \\
\multicolumn{1}{c}{mini-GOP 16} &  2.9 & 2.1  \\
\bottomrule
\end{tabular}
\label{tab:supp_Mini_GOP_Period}
\end{table}

%% file: table/Supp_Training_Details/Training_Procedures.tex
\begin{table*}[t]
\centering
\caption{Training procedure. MENet, MWNet, MCNet represent the motion estimation network, the motion extrapolation network, and the motion compensation network, respectively. $R^{motion}$ and $R_t$ represent the motion and total bitrates, respectively. EPA is error propagation aware training.}
\begin{tabular}{c ccccc}
\toprule
Phase & \begin{tabular}[c]{@{}c@{}}Number of\\ Frames\end{tabular} & \begin{tabular}[c]{@{}c@{}}Training \\ Modules\end{tabular} & Loss & lr & Epoch \\
\midrule
ME Training & 2 & MENet   & $D(x_t, \text{warp}(x_{t-1}, f_t))$  & 1e-5 & 5 \\
\cmidrule{2-6}
Motion Coding &3 & MWNet \& Motion codec   & $R^{\text{motion}} + \lambda \times D(x_t, \text{warp}(x_{t-1}, \hat{f}_t))$  & 1e-4 & 10 \\
\cmidrule{2-6}
MC Training &3 & MCNet                    & $R^{\text{motion}} + \lambda \times D(x_t, \hat{x}_C)$                       & 1e-4 & 3  \\

\cmidrule{2-6}
\multirow{3}{*}{\begin{tabular}[c]{@{}c@{}}Inter-Frame Coding\end{tabular}} &3 & Inter codec & $R_t + \lambda \times D(x_t, \hat{x}_t)$ & 1e-4 & 10 \\
 &5 & Inter codec  & $R_t + \lambda \times D(x_t, \hat{x}_t)$ & 1e-4 & 2  \\
&5 & Inter codec  \& MCNet & $R_t + \lambda \times D(x_t, \hat{x}_t)$ & 1e-4 & 7 \\
\cmidrule{2-6}
Finetune &5 & All modules except MENet & $R_t + \lambda \times D(x_t, \hat{x}_t)$ & 1e-4 & 5  \\
\cmidrule{2-6}
\multirow{2}{*}{\begin{tabular}[c]{@{}c@{}}Finetune + EPA \end{tabular}} &5 & All modules except MENet & $R_t + \lambda \times  D(x_t, \hat{x}_t)$ & 1e-5 & 1  \\
&5 & All modules except MENet & $R_t + \lambda \times w_t \times D(x_t, \hat{x}_t)$ & 1e-5 & 2  \\
\midrule
Variable Rate Finetune
&5 & All modules  except MENet    & $R_t + \lambda_p \times w_t \times D(x_t, \hat{x}_t)$ & 1e-5 & 3  \\

\bottomrule
\end{tabular}
\label{tab:supp_training_procedure}
\end{table*}

%% file: section/G_Coding_Structure.tex
\input{FIG_TXT/Architectures/4-1_Pred_Struct}

\section{Cascading and Prediction Errors}
\label{sec:first}

\input{FIG_TXT/Per_Frame/Cascading_Error}

\input{FIG_TXT/Per_Frame/Prediction_Error}

Fig.~\ref{fig:supp_cascading_error} shows that our $LS$ is more effective than $TP$ in mitigating temporal cascading errors. The results stress the importance of incorporating both long- and short-term reference frames.

Fig.~\ref{fig:supp_prediction_error} further evaluates the prediction errors of $LS$ and $LL$ by visualizing the PSNR-RGB of the temporal predictor $x_c$. These results are evaluated for fast-motion sequences, where the prediction errors are more noticeable. $LL$ exhibits inferior temporal predictor quality compared to $LS$. Notably, the quality of the temporal predictor with $LL$ degrades over time within a mini-GOP due to the increasing prediction distance between the coding frame and its reference frames. In contrast, the temporal predictor quality of $LS$ remains relatively stable over time, as it leverages both long- and short-term reference frames.

%% file: FIG_TXT/Architectures/4-1_Pred_Struct.tex
\begin{figure}[t!]
    \centerline{\includegraphics[width=0.48\textwidth]{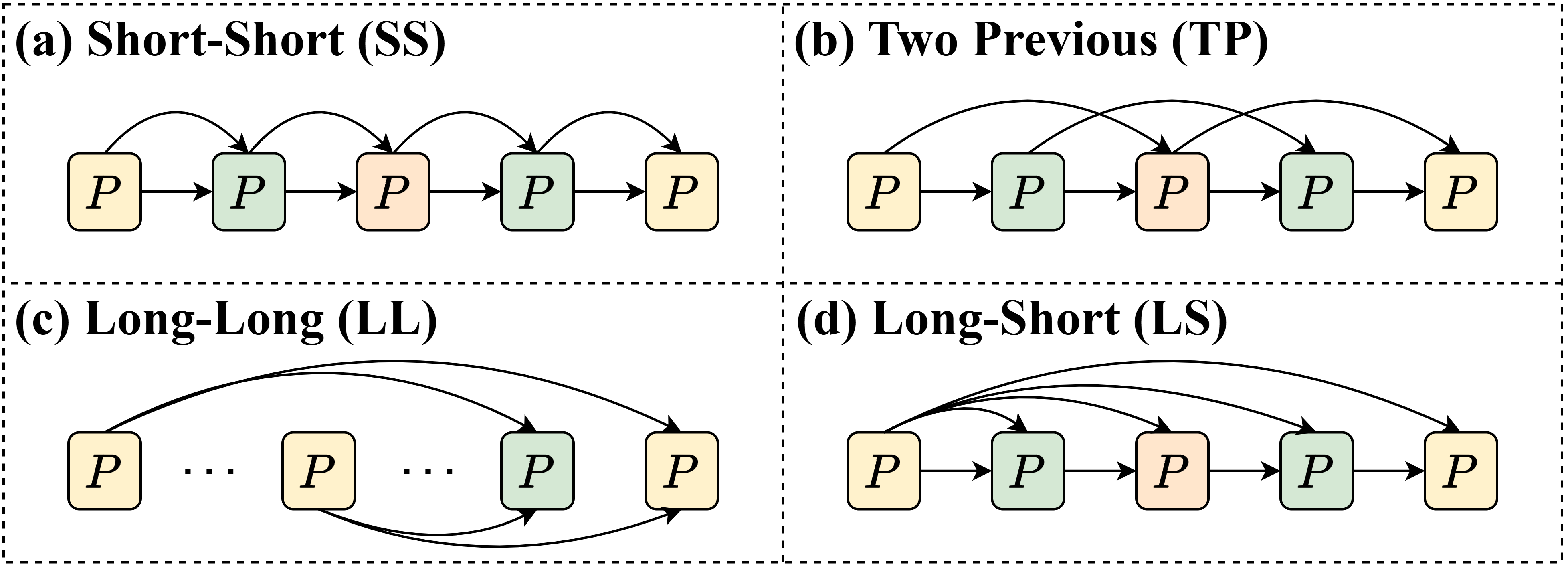}}
    \caption{Alternative prediction structures.}
    \label{fig:supp_pred_struct} 
    \vspace{-1em}
    \label{fig:supp_alter_pred}
\end{figure}

%% file: FIG_TXT/Per_Frame/Cascading_Error.tex
\begin{figure}[t]
    \centerline{\includegraphics[width=0.48\textwidth]{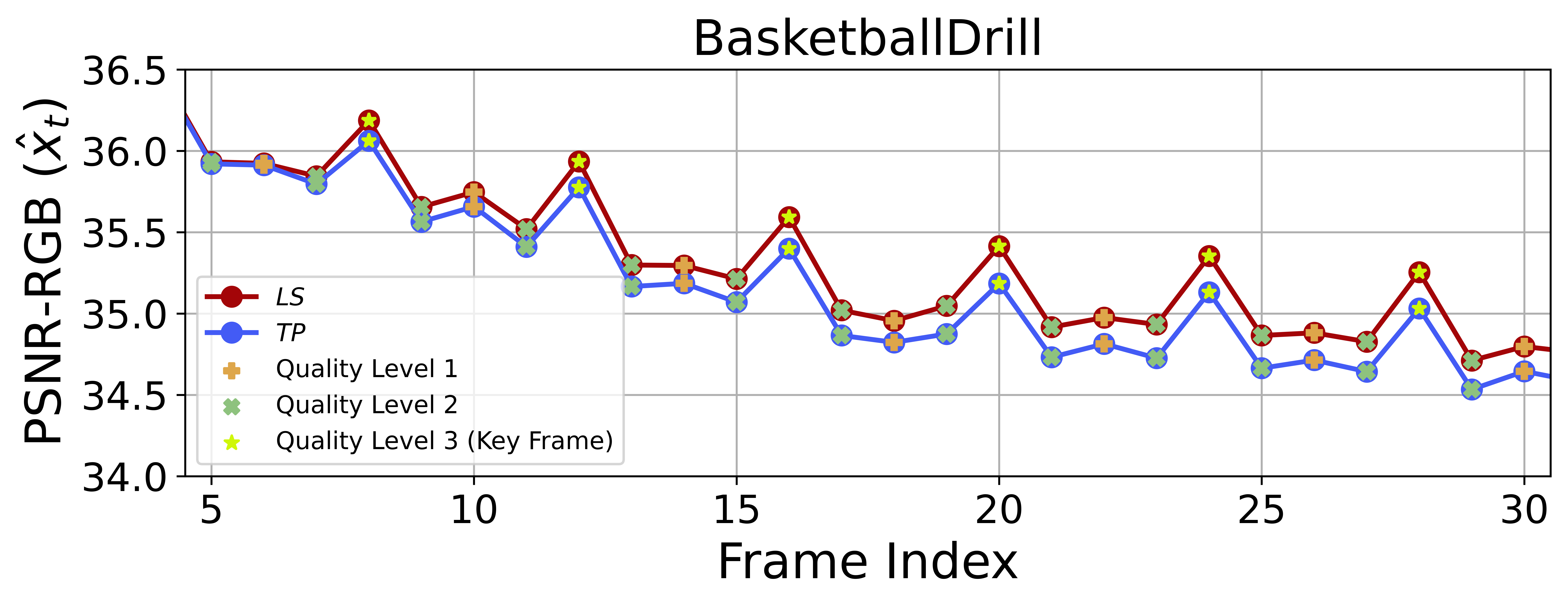}}
    \centerline{\includegraphics[width=0.48\textwidth]{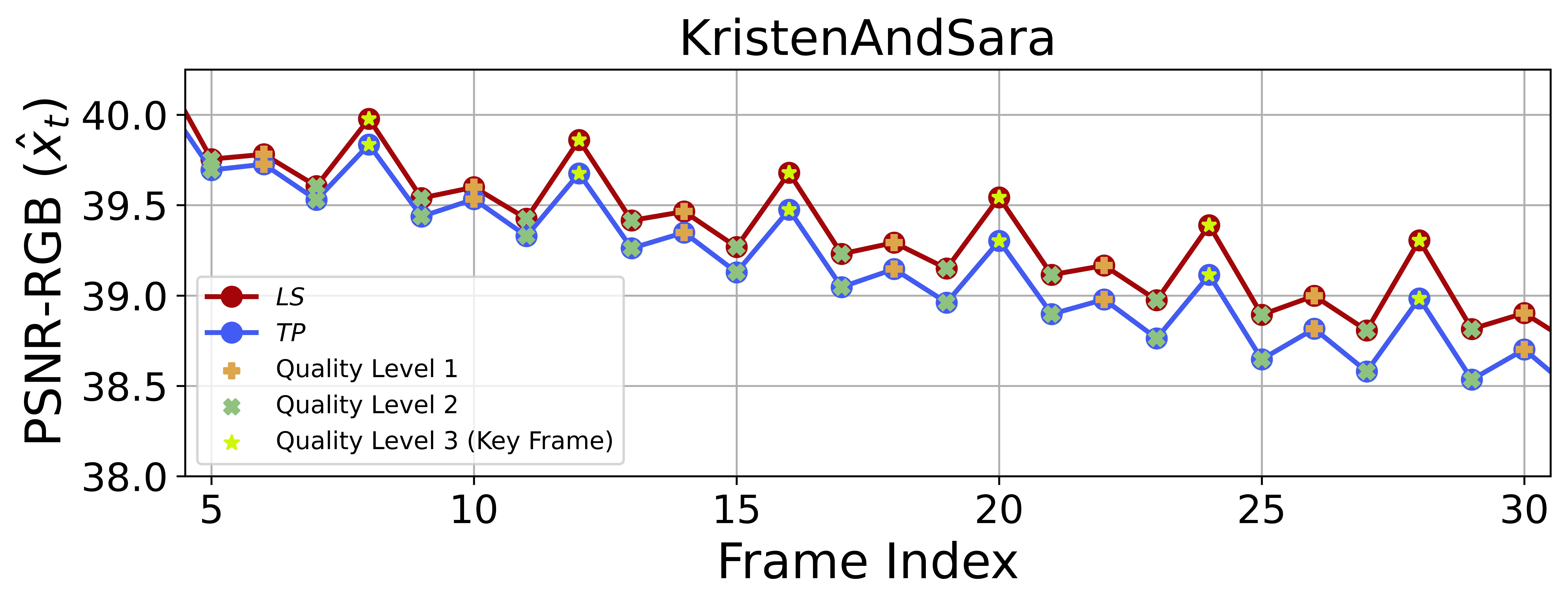}}
    \caption{The per-frame PSNR profiles on BasketballDrill and KristenAndSara: $LS$ versus $TP$. Their average bitrates are comparable.}
    \label{fig:supp_cascading_error}
\end{figure}

%% file: FIG_TXT/Per_Frame/Prediction_Error.tex
\begin{figure}[t]
    \centerline{\includegraphics[width=0.48\textwidth]{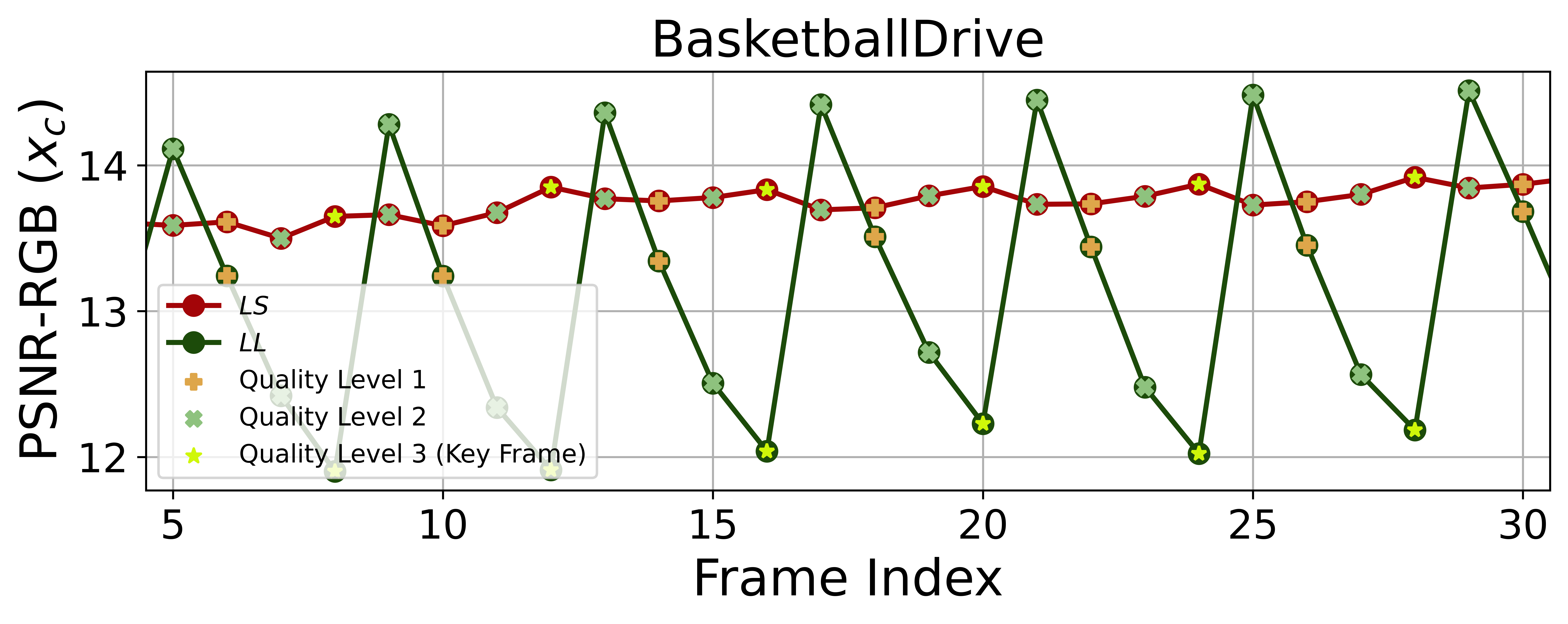}}
    \centerline{\includegraphics[width=0.48\textwidth]{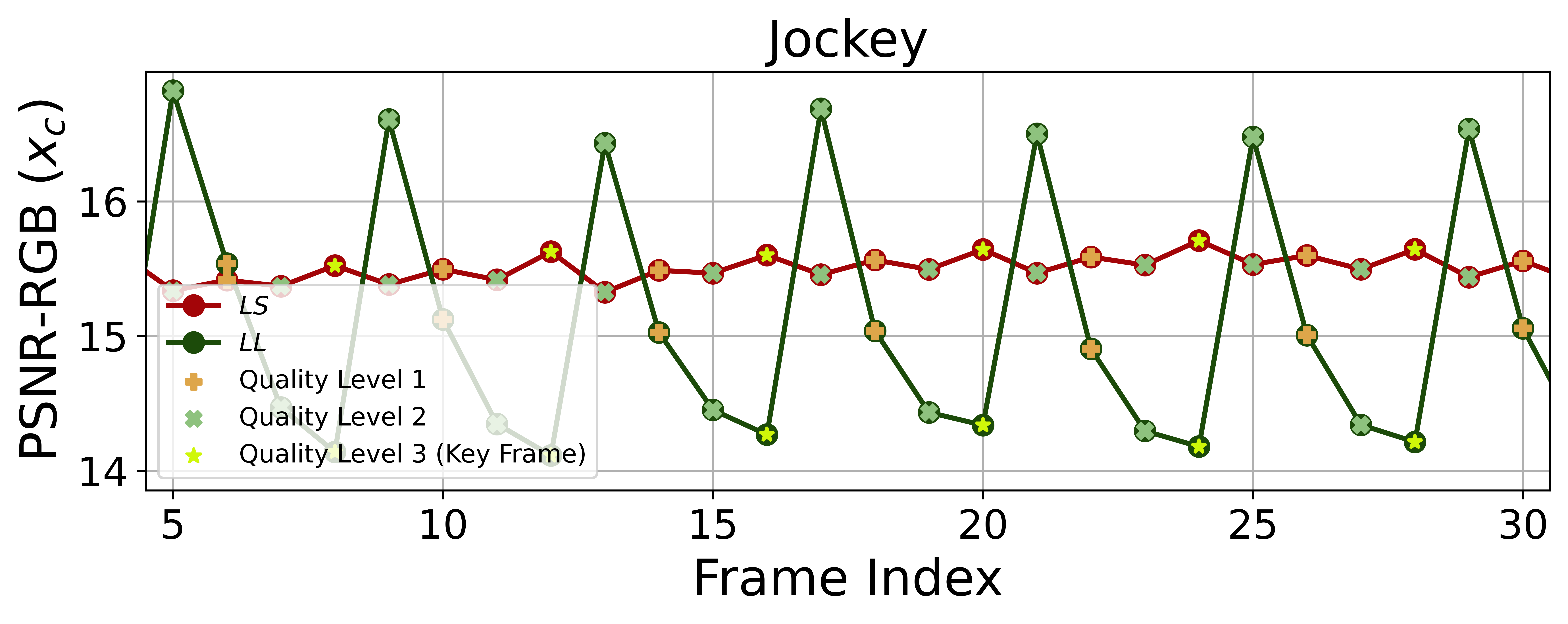}}
    \caption{The per-frame PSNR profiles of the temporal predictor $x_c$ on BasketballDrive and Jockey: $LS$ versus $LL$. Their average bitrates are comparable. } 
    \label{fig:supp_prediction_error}
\end{figure}

%% file: section/E_TCM_overview.tex
\section{Overview of Implicit Buffering Strategies.}
\label{sec:implicit}
\input{FIG_TXT/Architectures/TCM_overview}

Figure~\ref{fig:supp_TCM_overview} provides an overview of the implicit buffering strategies framework. To ensure a fair comparison, we implement the implicit buffering strategy of DCVC-TCM~\cite{tcmli} within the same conditional residual video coding framework as our proposed MH-CRT (see Section ~\ref{subsubsec:implicit_buffering_comparison} in the main paper).

%% file: FIG_TXT/Architectures/TCM_overview.tex
\begin{figure*}[t]
    \centerline{\includegraphics[width=1\textwidth]{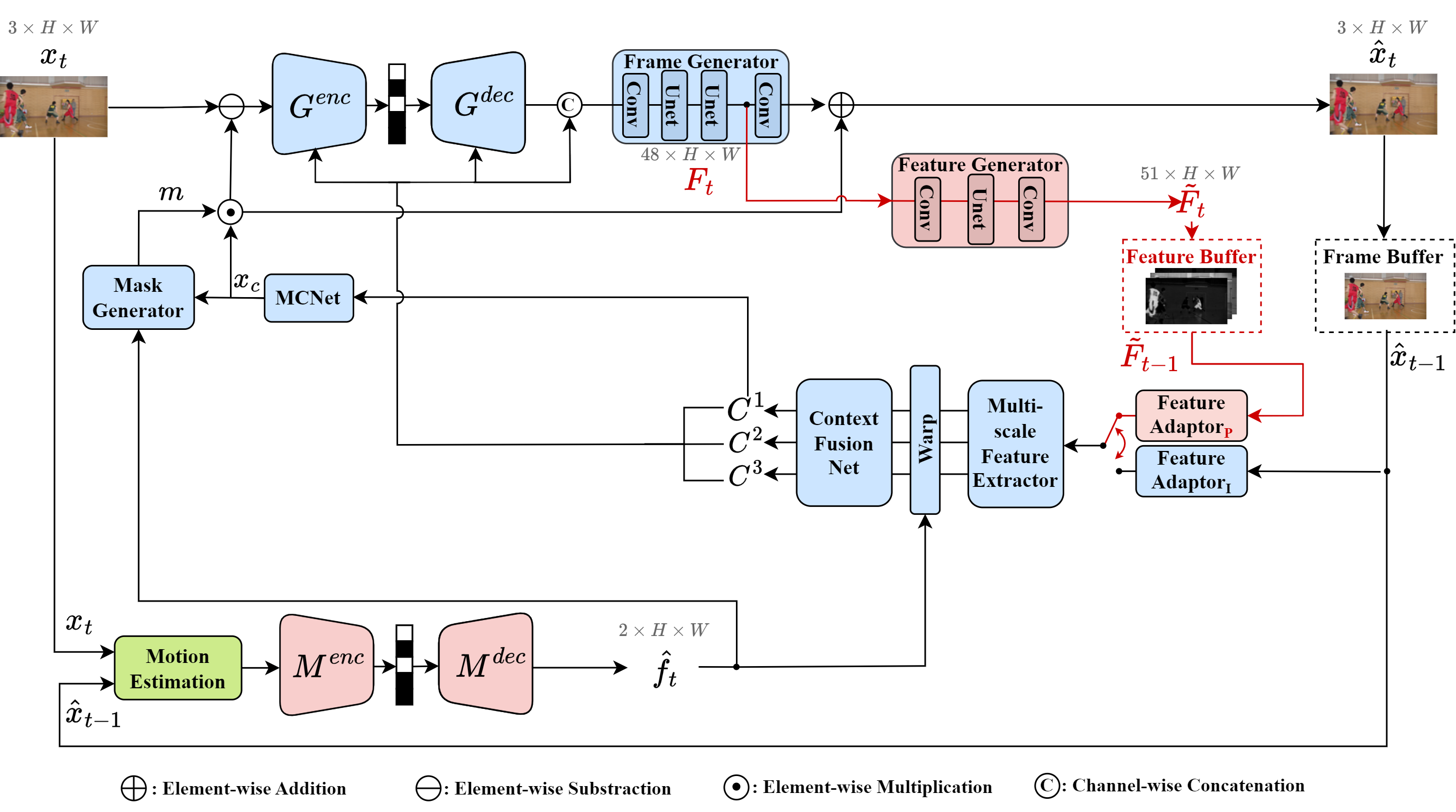}}
    \vspace{-1.0em}
    \caption{Overview of the implicit buffering strategies framework.}
    \label{fig:supp_TCM_overview}
\end{figure*}

%% file: section/C_Alternative_pred.tex
\section{Additional Alternative Temporal Prediction Structures} 
\label{sec:alter_pred}

\input{table/Supp_Rebuttal/B_All_Pred_Structure}

We further investigate several alternative prediction structures (see Fig.~\ref{fig:supp_pred_struct}) to prove the effectiveness of $LS$. As mentioned in the main paper, (i) ``Short-Short ($SS$), which simulates the effect of the single-hypothesis prediction with our two-hypothesis framework by referencing the most recently decoded frame twice yet with the same optical flow 
map, (ii) ``Two Previous ($TP$),'' which predicts a current frame from the last two previously decoded frames.  We also include ``Two Previous Plus ($TP^{+}$),'' which uses the most recently decoded frame $\hat{x}_{t-1}$ as the short-term reference frame and selects adaptively another short-term reference frame from $\hat{x}_{t-2}$, $\hat{x}_{t-3}$, (iv) “Long-Long (LL),” which predicts from the last two long-term key frames.The ``Long-Short ($LS$)'' corresponds to our MH-LVC-1 prediction scheme, which has one short-term reference frame and one long-term key frame. Following the same notation as $TP^{+}$, we use $LS^{+}$ to denote our MH-LVC-2, which has two long-term key frames for adaptive prediction. Note that all the prediction structures share the same network weights trained solely for the $LS$ prediction. Table \ref{tab:supp_diff_pred} justifies the effectiveness of $LS^{+}$ over $SS$, $TP$ and $TP^{+}$. 

%% file: table/Supp_Rebuttal/B_All_Pred_Structure.tex
\begin{table}[h!]
\centering
\caption{BD-rate comparison of several prediction structures with $LS^+$ serving as the anchor.}

\label{tab:content_specific}
\footnotesize
\begin{tabular}{c | cc  ccccc}
\midrule
       & $LS$ & $LS^+$ & $SS$ & $TP$ & $TP^+$  \\
\midrule
  HEVC-B     & 0.0  & \textbf{-3.4} & 14.6 & 3.3 & -0.8 \\
 UVG        & 0.0  & \textbf{-2.7} & 10.5 & 2.4 & -0.5 \\
 MCL-JCV    & 0.0  & \textbf{-4.6} & 10.5 & 2.4 & -3.2 \\
\midrule
\end{tabular}

\label{tab:supp_diff_pred}
\end{table}

%% file: section/I_MRF.tex
\section{The Number of Long-term Key Frames}
\label{sec:mrf}

\input{table/Supp_Ablations/MRF/Multiple_Key}
Table~\ref{tab:supp_key_num} presents how the number of long-term key frames under our LS prediction structure may impact the complexity and compression performance. It is seen that the coding gain diminishes when the number of long-term key frames goes beyond 2, while the encoding kMAC/pixel increases considerably. 

%% file: table/Supp_Ablations/MRF/Multiple_Key.tex
\begin{table}[t]
\centering
\footnotesize
\caption{Ablation on the number of key frames for online long-term key frame selection. The anchor is MH-LVC-1.}
\begin{tabular}{ccc cc}
\toprule

\multirow{2}{*}{Number of} & \multirow{2}{*}{Encoding}  & \multicolumn{2}{c}{BD-rate (\%) PSNR-RGB} \\

\cmidrule(r){3-4}
Key Frames & KMACs/pixel & HEVC-B & UVG \\
\midrule
1 & 1507 & \;  0.0 & \; 0.0 \\
2 & 2611 & -3.4 & -2.7 \\
3 & 3716 & -3.8 & -3.0 \\
\bottomrule

\end{tabular}
\label{tab:supp_key_num}
\vspace{-1em} 
\end{table}

%% file: section/H_Traditional_Codecs.tex
\section{Command Lines for VTM and HM}
\label{sec:traditional}
We compare MH-LVC with traditional video codecs, including VTM-17.0 and HM-16.25. Following~\cite{dcvcdc}, we have these codecs encode input videos in YUV444 format (by converting them from YUV420 into YUV444). The reconstructed YUV444 videos are then transformed into RGB domain for evaluating the distortions. For HM and VTM, \emph{encoder\_lowdelay\_main\_rext.cfg} and \emph{encoder\_lowdelay\_vtm.cfg} config files are used, respectively. The command lines used are as follows:
\vspace{2em} 

\begin{itemize}[label=\textbullet]
    \item \texttt{-c \{config file name\}}\\
    \texttt{--InputFile=\{input video name\}}\\
    \texttt{--InputBitDepth=\{input bit depth\}}\\
    \texttt{--OutputBitDepth=\{output bit depth\}}\\
    \texttt{--InputChromaFormat=444}\\
    \texttt{--FrameRate=\{frame rate\}}\\
    \texttt{--DecodingRefreshType=\{refresh type\}}\\
    \texttt{--FramesToBeEncoded=96}\\
    \texttt{--SourceWidth=\{width\}}\\
    \texttt{--SourceHeight=\{height\}}\\
    \texttt{--IntraPeriod=\{intra period\}}\\
    \texttt{--QP=\{qp\}}\\
    \texttt{--BitstreamFile=\{bitstream file name\}}\\
\end{itemize}
where we set \emph{DecodingRefreshType} and \emph{IntraPeriod} to 2 and 32, respectively, for an intra period of 32. They are set to 0 and -1, respectively, for an infinite intra period.

%% file: section/J_Compare_SOTA.tex
\section{Comparison with the State-of-the-art Methods in Terms of MS-SSIM-RGB}
\label{sec:com_SOTA}

Fig.~\ref{fig:supp_SSIM_I32} and Fig.~\ref{fig:supp_SSIM_INF} show the rate-distortion curves of our MH-LVC under an intra-period of 32 and infinity, respectively, where the quality metric is MS-SSIM-RGB.

\input{table/Supp_RD_Plot/SSIM_I32}
\input{table/Supp_RD_Plot/SSIM_INF}

\input{table/Supp_RD_Plot/RGB_I32}
\input{table/Supp_RD_Plot/RGB_INF}

%% file: table/Supp_RD_Plot/SSIM_I32.tex
\begin{figure*}[tbp]
    \begin{center}
    \begin{subfigure}{0.49\linewidth}
        \centering
        \includegraphics[width=\linewidth]{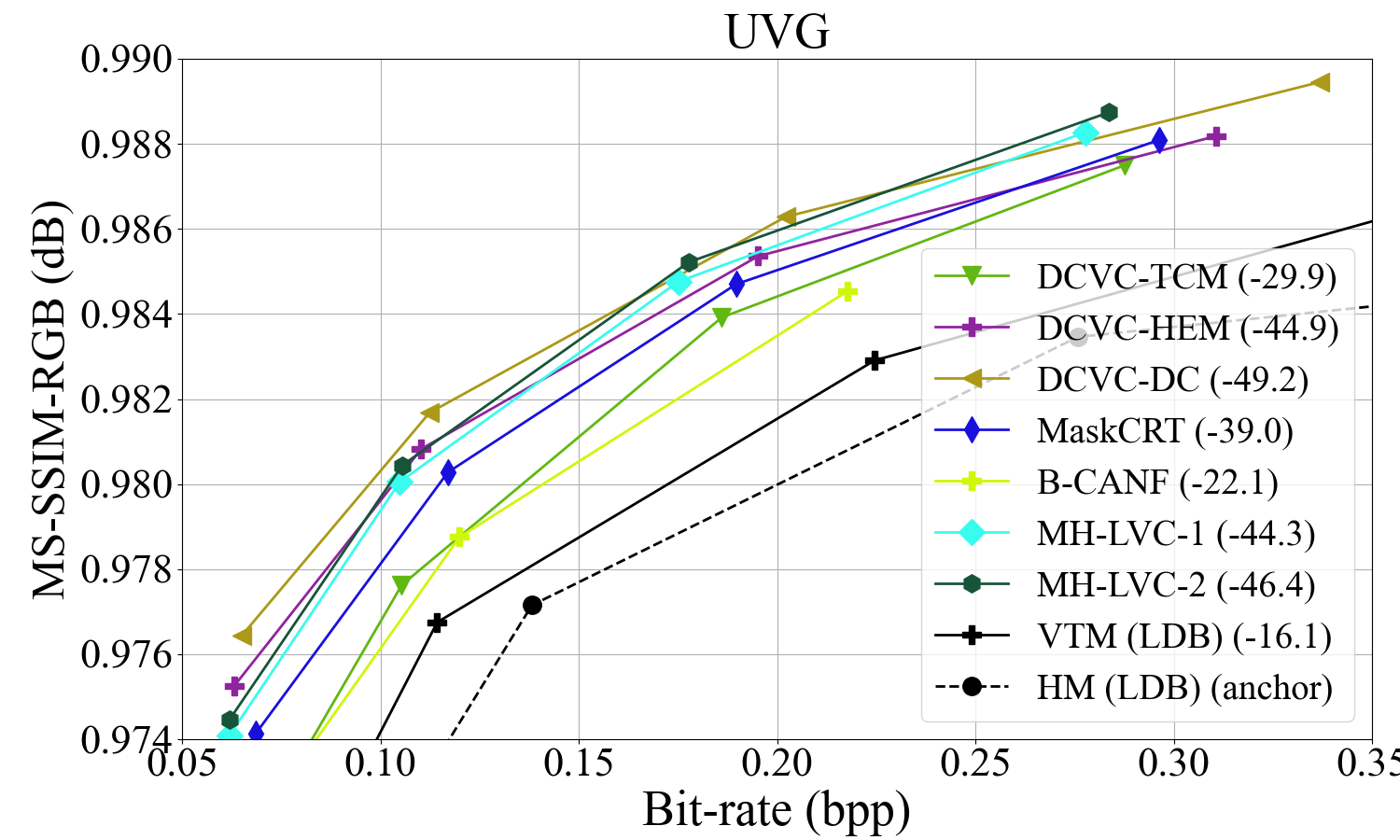}
    \end{subfigure}
    \begin{subfigure}{0.49\linewidth}
        \centering
        \includegraphics[width=\linewidth]{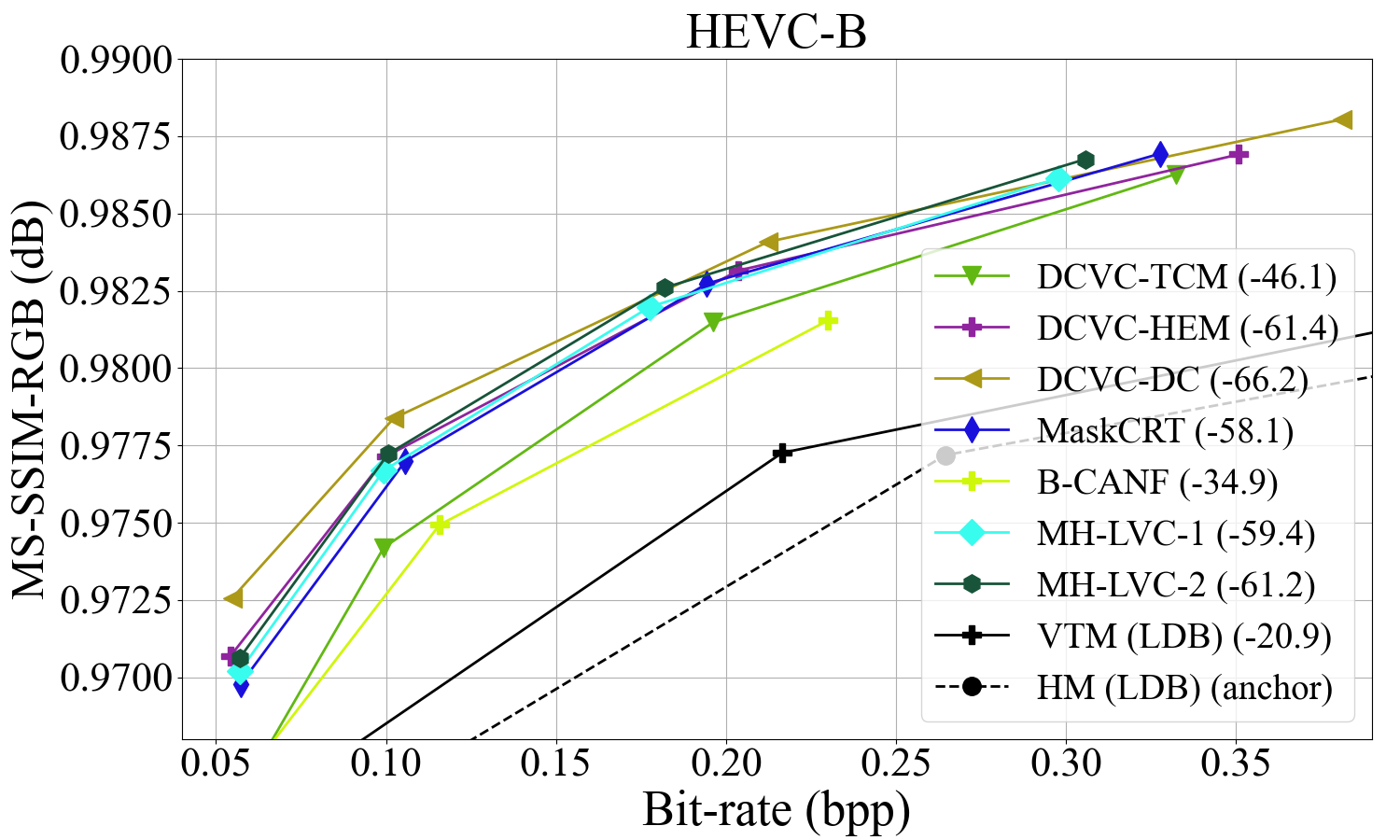}
    \end{subfigure}
    \begin{subfigure}{0.49\linewidth}
        \centering
        \includegraphics[width=\linewidth]{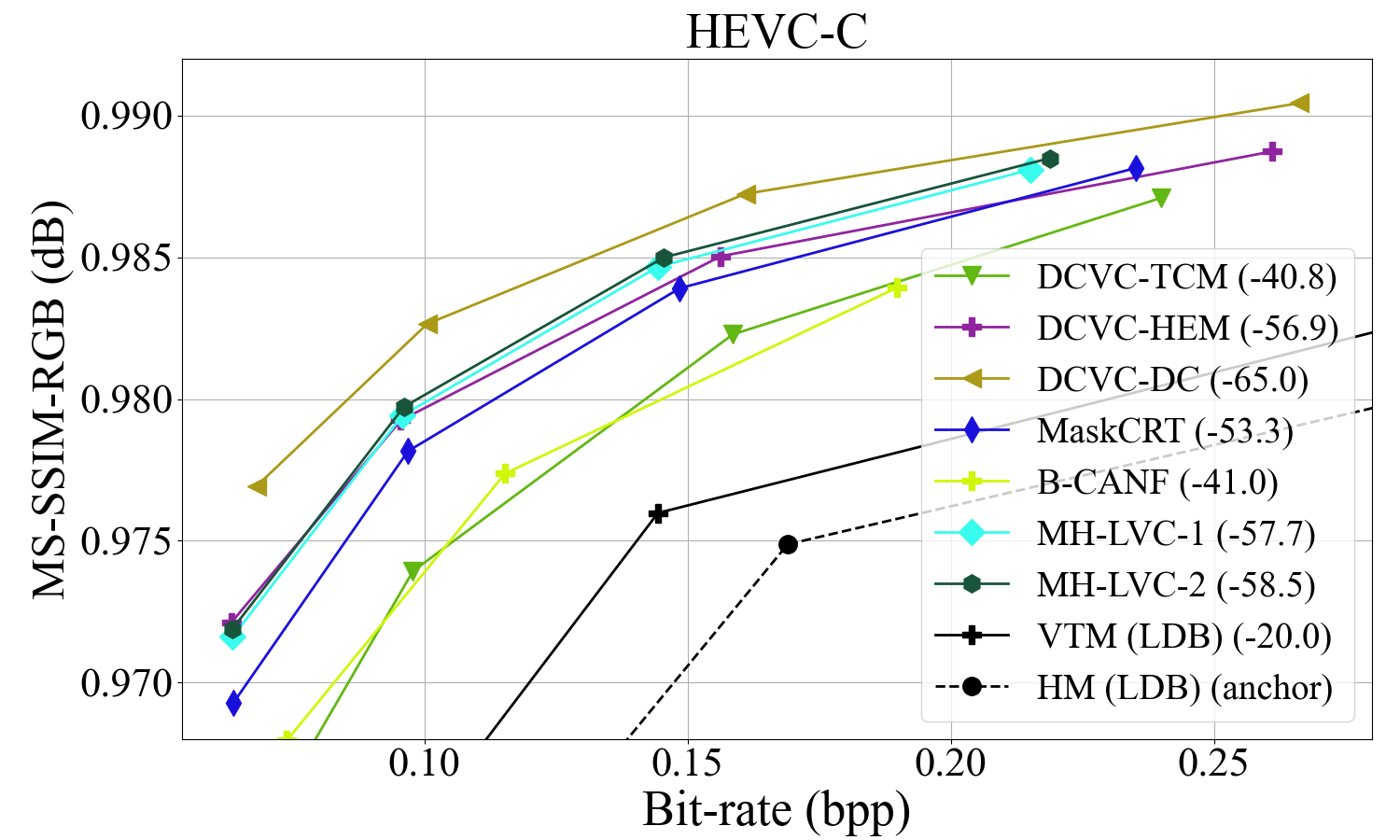}
    \end{subfigure}
    \begin{subfigure}{0.49\linewidth}
        \centering
        \includegraphics[width=\linewidth]{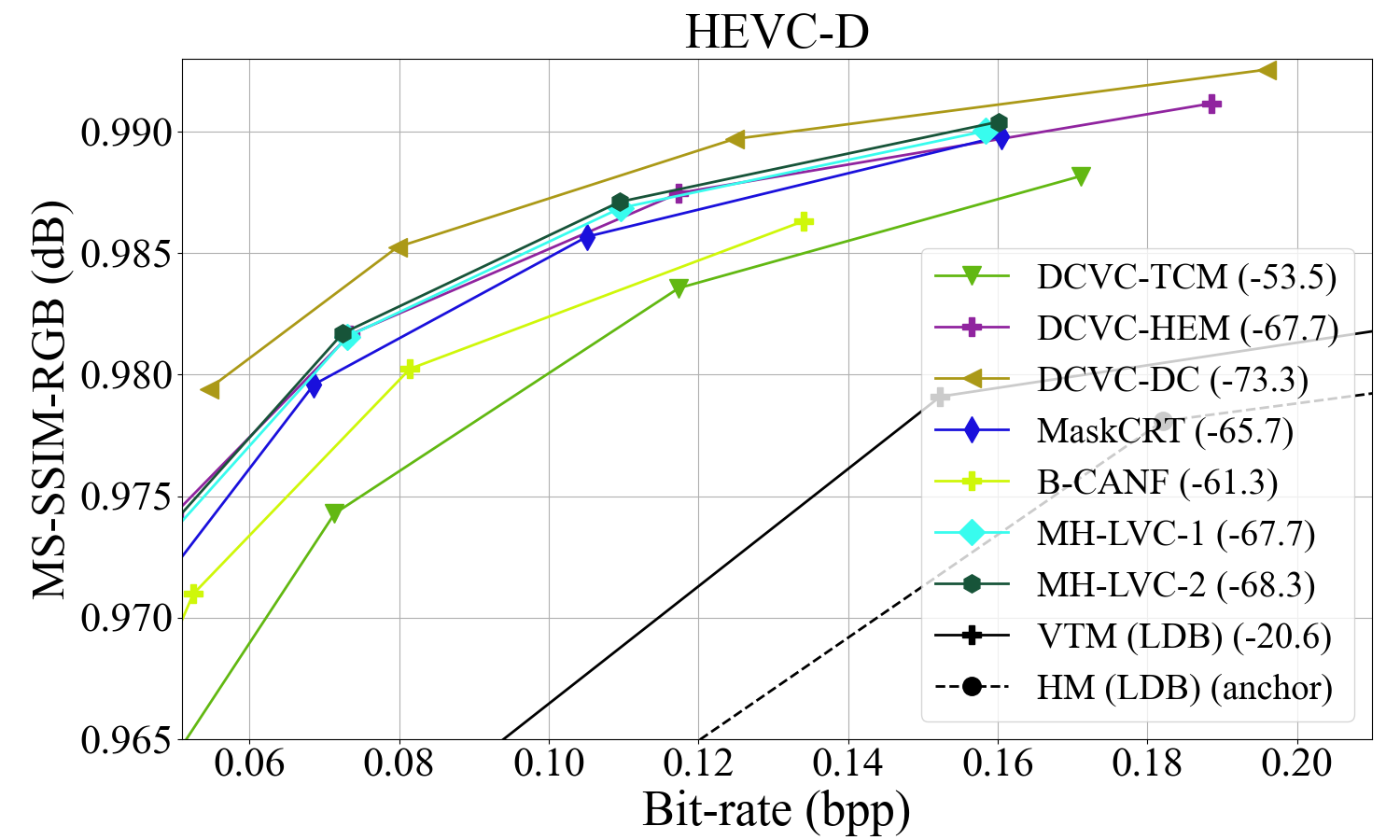}
    \end{subfigure}
    \begin{subfigure}{0.49\linewidth}
        \centering
        \includegraphics[width=\linewidth]{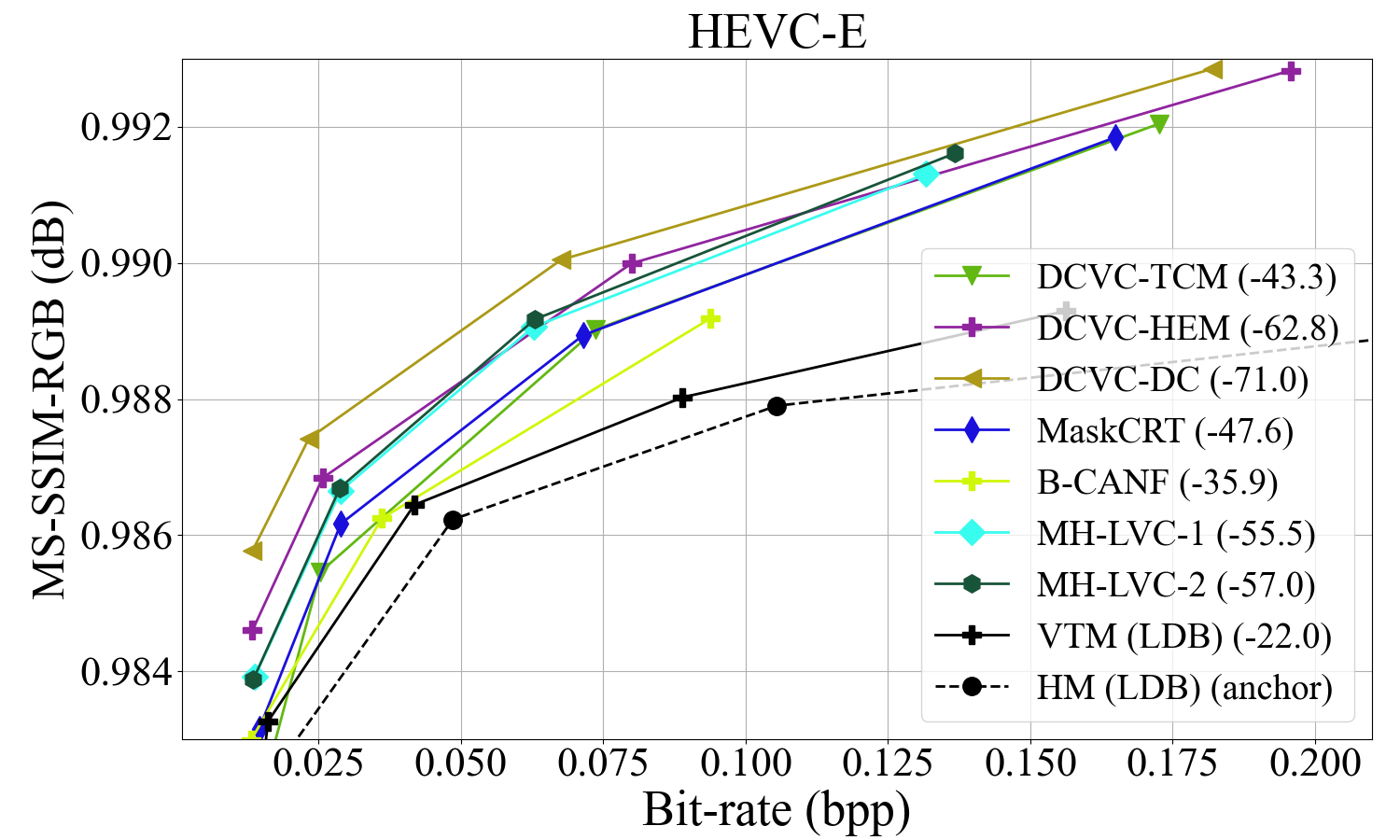}
    \end{subfigure}
    \begin{subfigure}{0.49\linewidth}
        \centering
        \includegraphics[width=\linewidth]{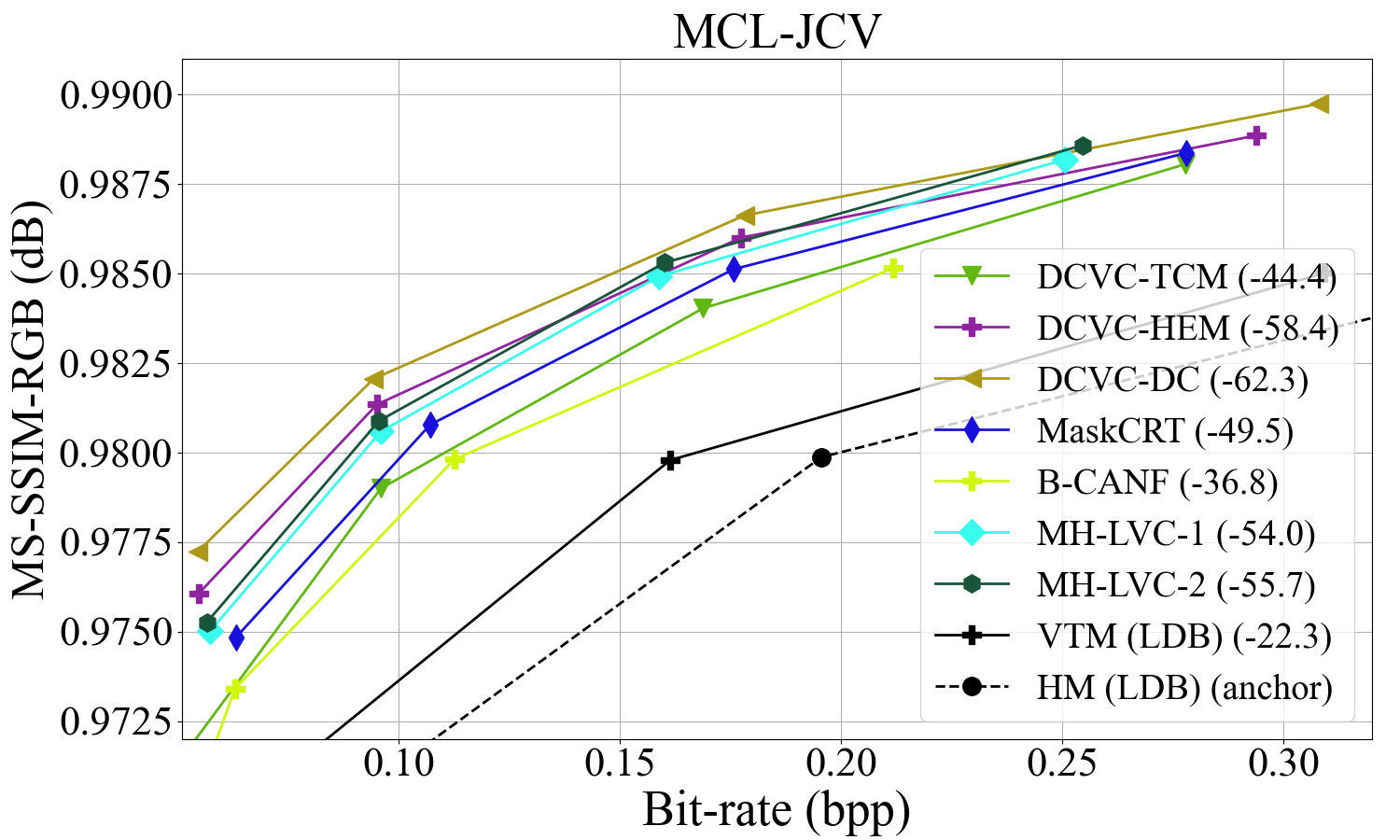}
    \end{subfigure}
    \caption{Rate-distortion performance comparison with intra-period 32 in terms of MS-SSIM-RGB. The numbers within the parentheses are BD-rates, with HM-16.25 (Low-delay B) serving as the anchor.}
    \label{fig:supp_SSIM_I32}
    \end{center}
\end{figure*}

%% file: table/Supp_RD_Plot/SSIM_INF.tex
\begin{figure*}[tbp]
    \begin{center}
    \begin{subfigure}{0.49\linewidth}
        \centering
        \includegraphics[width=\linewidth]{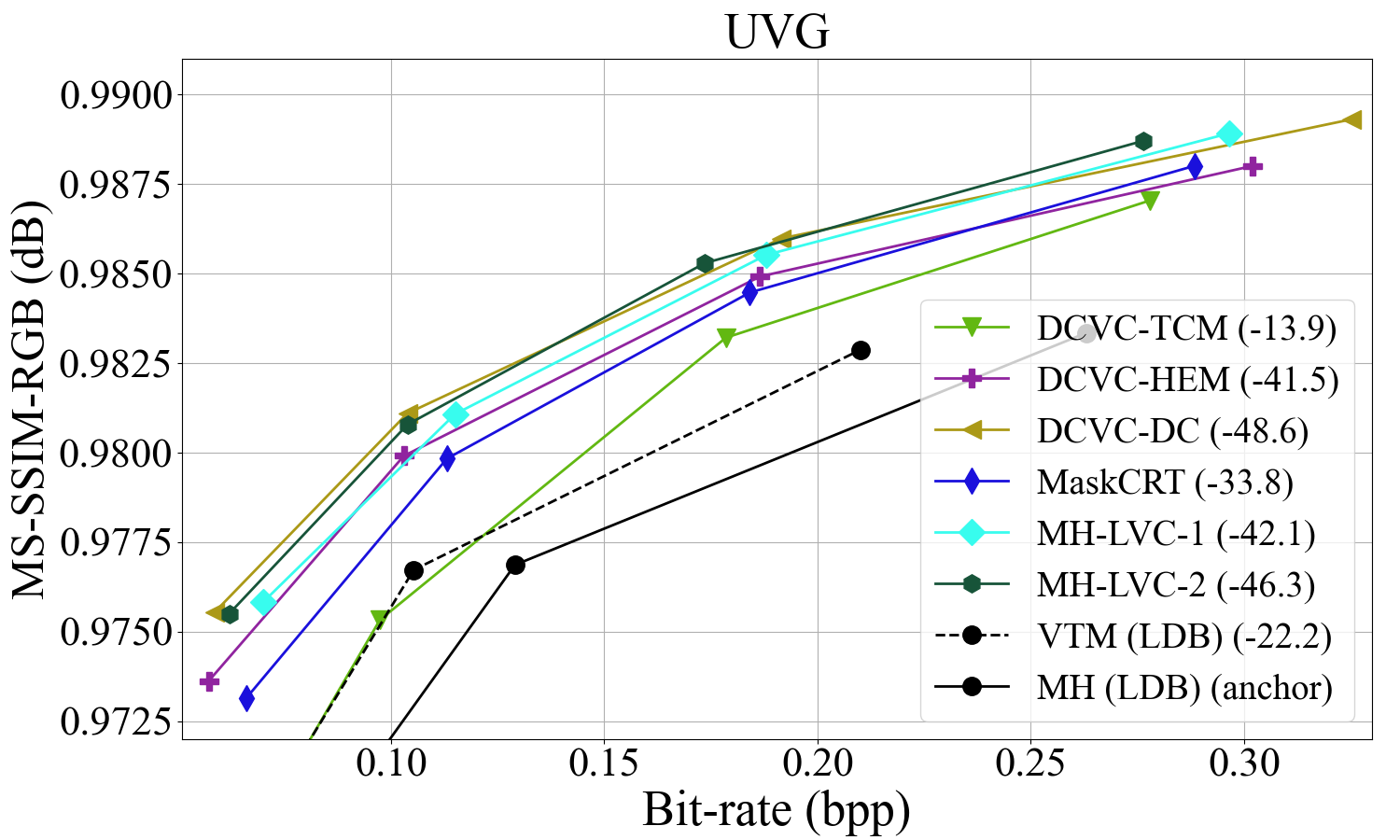}
    \end{subfigure}
    \begin{subfigure}{0.49\linewidth}
        \centering
        \includegraphics[width=\linewidth]{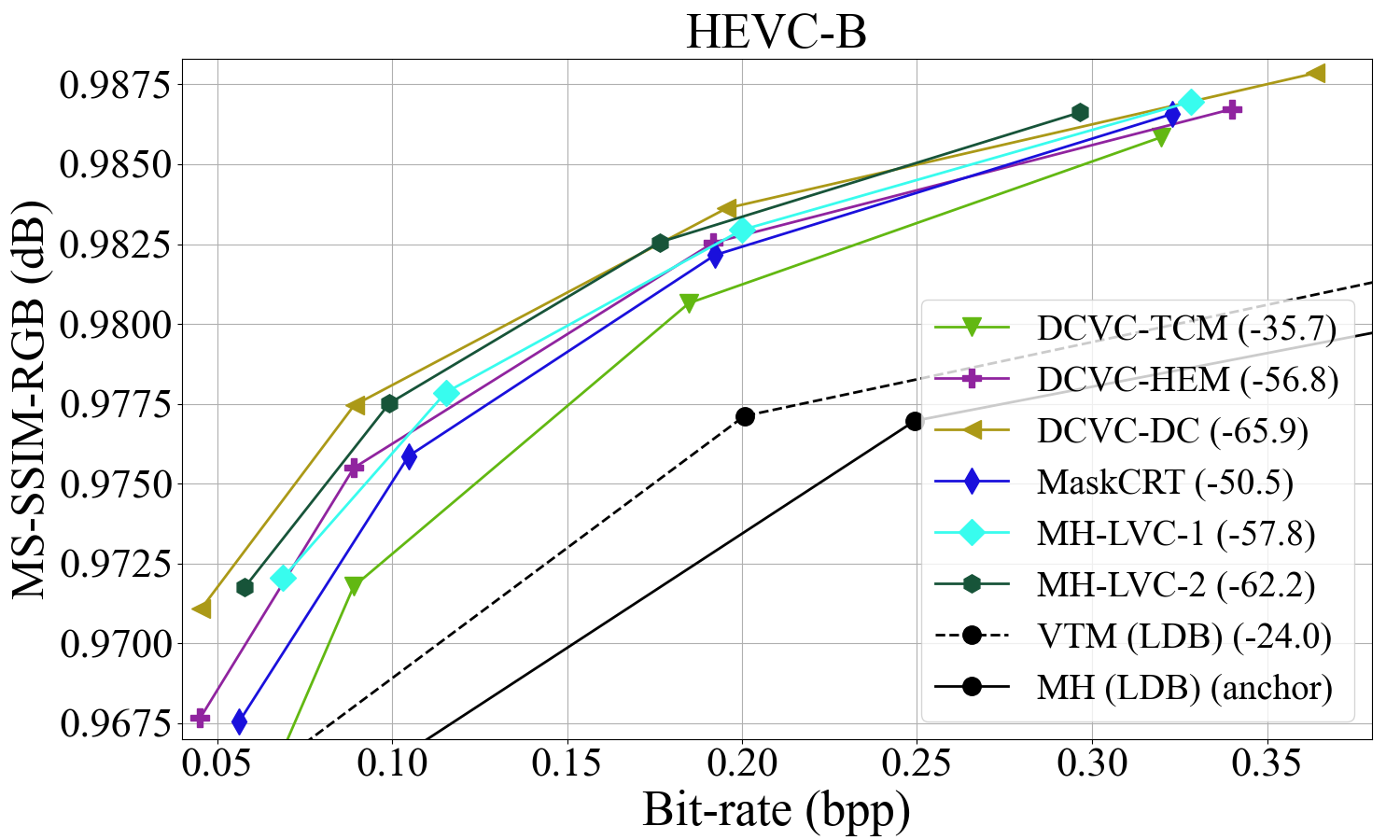}
    \end{subfigure}
    \begin{subfigure}{0.49\linewidth}
        \centering
        \includegraphics[width=\linewidth]{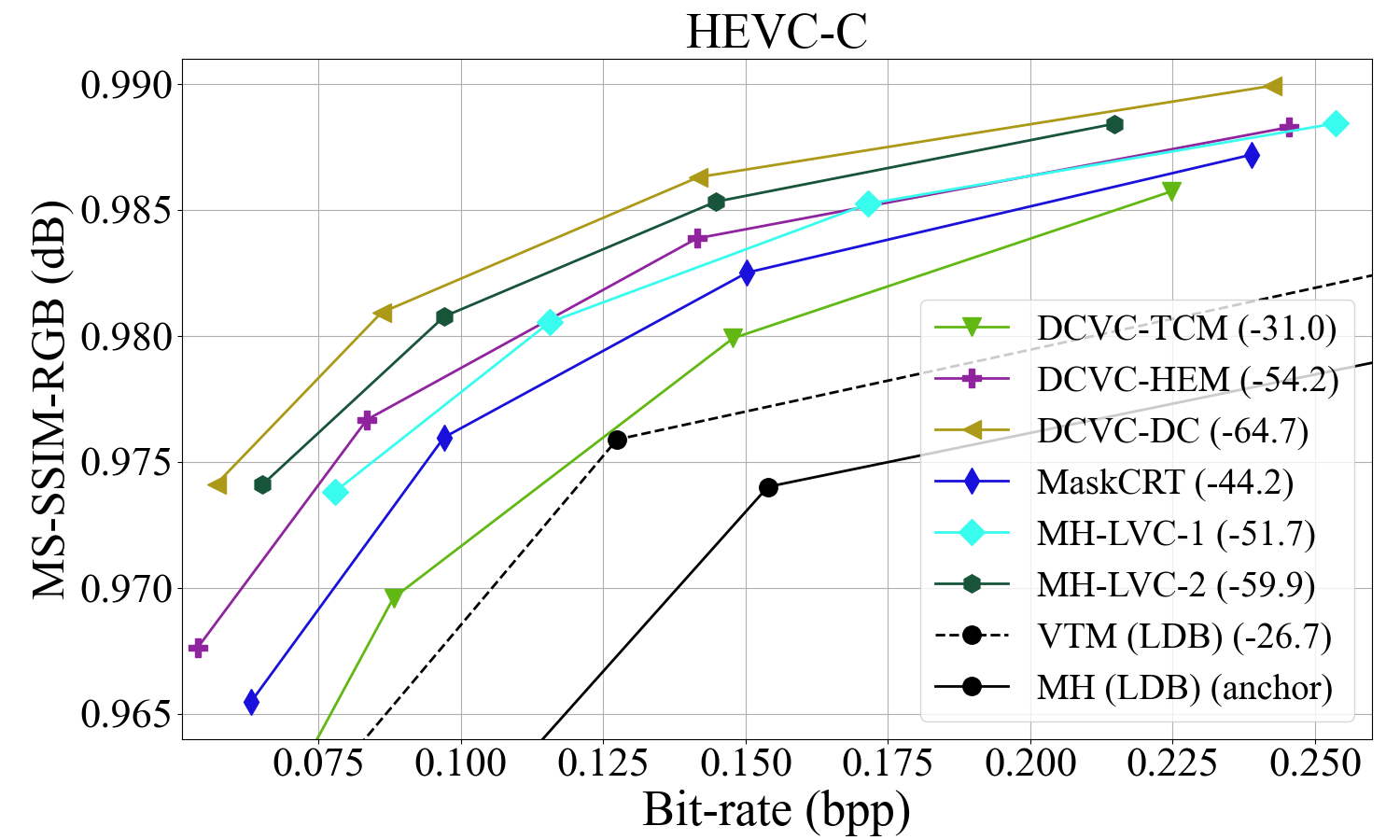}
    \end{subfigure}
    \begin{subfigure}{0.49\linewidth}
        \centering
        \includegraphics[width=\linewidth]{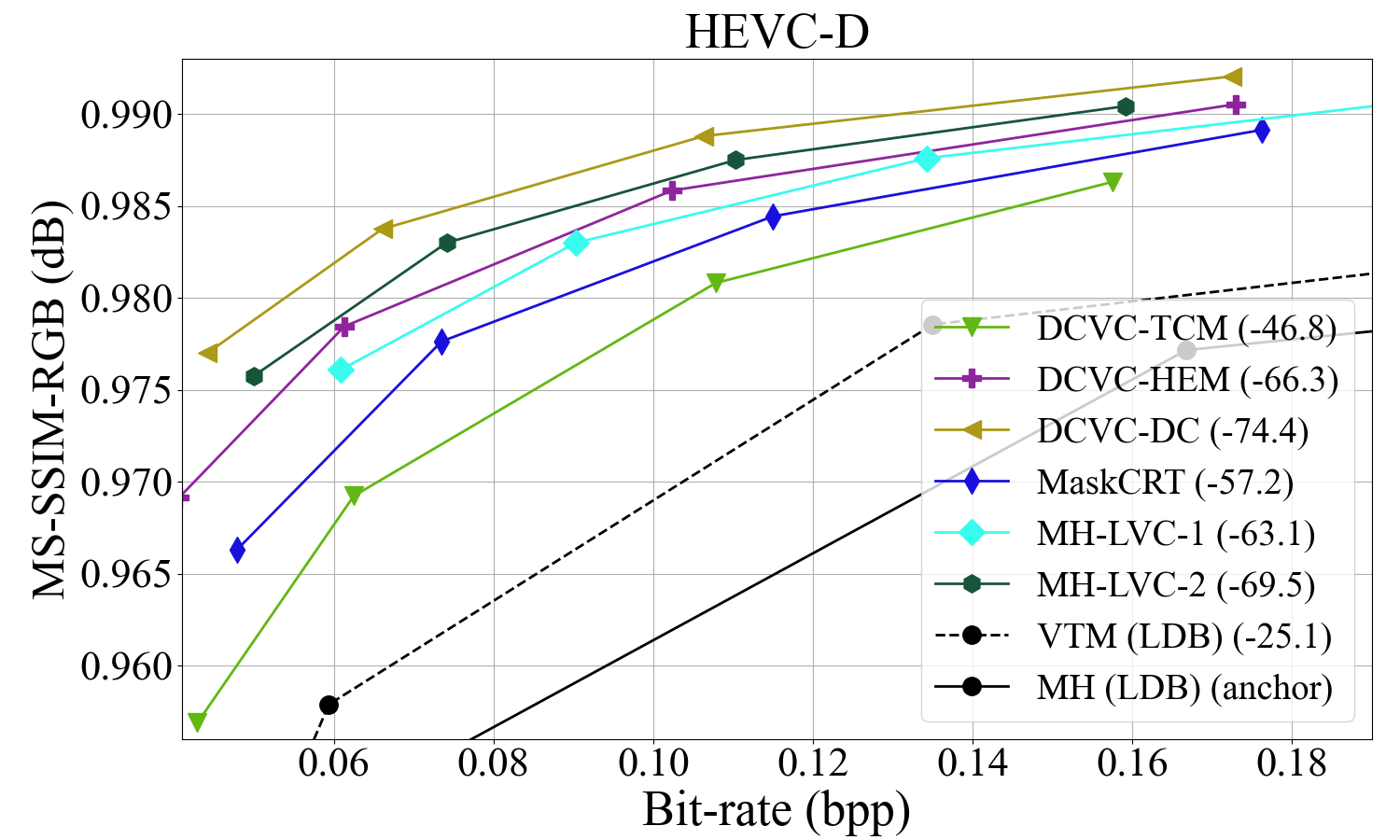}
    \end{subfigure}
    \begin{subfigure}{0.49\linewidth}
        \centering
        \includegraphics[width=\linewidth]{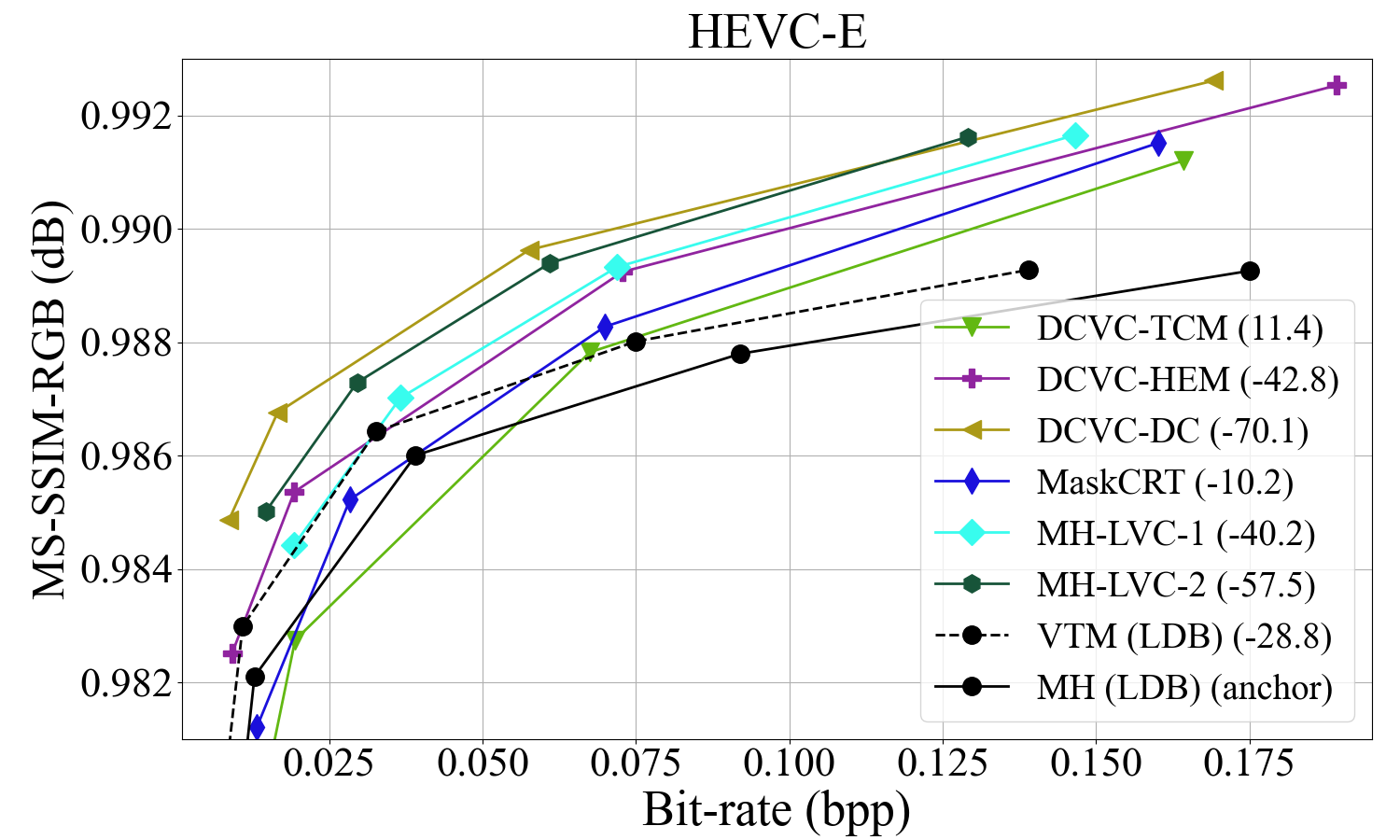}
    \end{subfigure}
    \begin{subfigure}{0.49\linewidth}
        \centering
        \includegraphics[width=\linewidth]{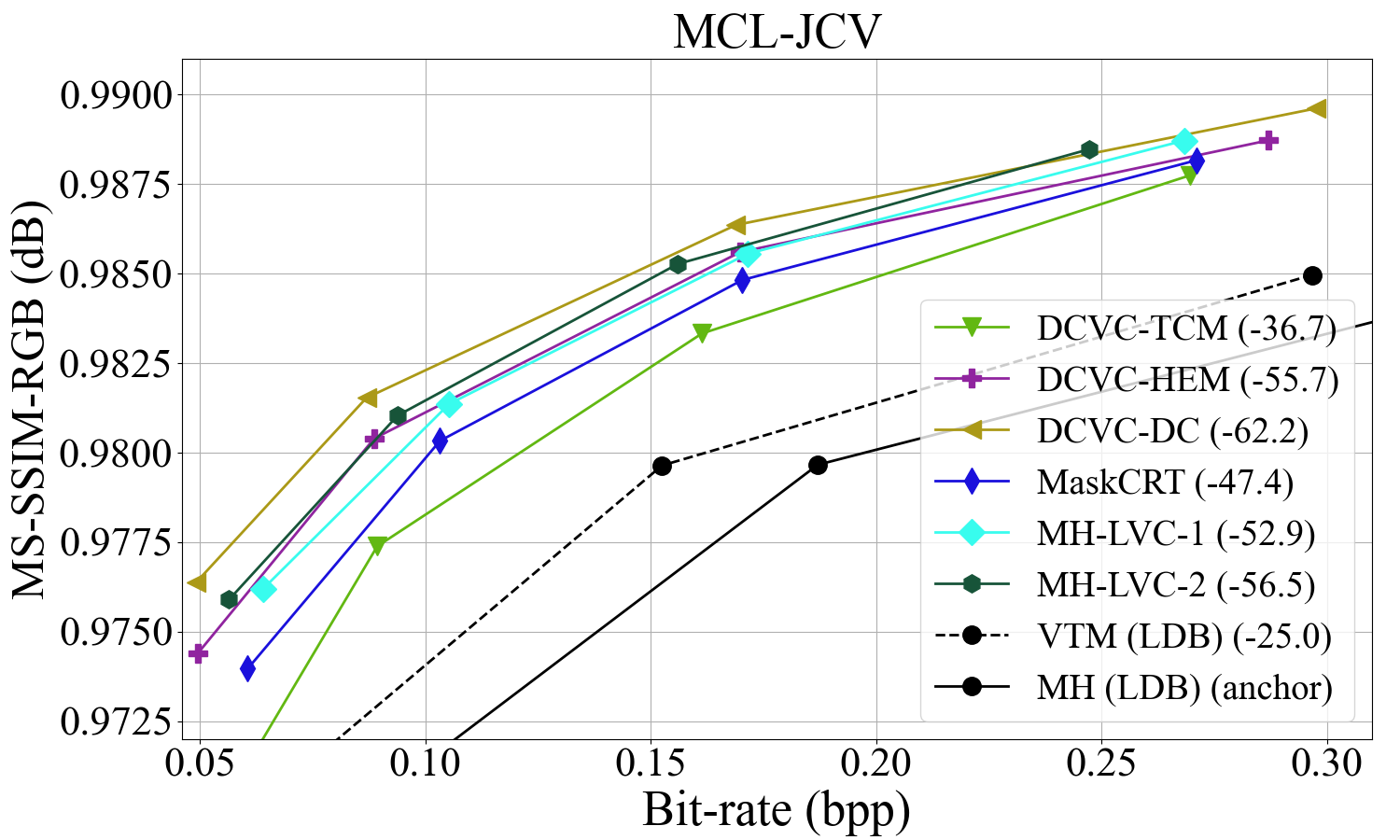}
    \end{subfigure}
    \caption{Rate-distortion performance comparison under an infinite intra-period in terms of MS-SSIM-RGB. The numbers within the parentheses are BD-rates, with HM-16.25 (Low-delay B) serving as the anchor.}
    \label{fig:supp_SSIM_INF}
    \end{center}
\end{figure*}

%% file: table/Supp_RD_Plot/RGB_I32.tex
\begin{figure*}[tbp]
    \begin{center}
    \begin{subfigure}{0.45\linewidth}
        \centering
        \includegraphics[width=\linewidth]{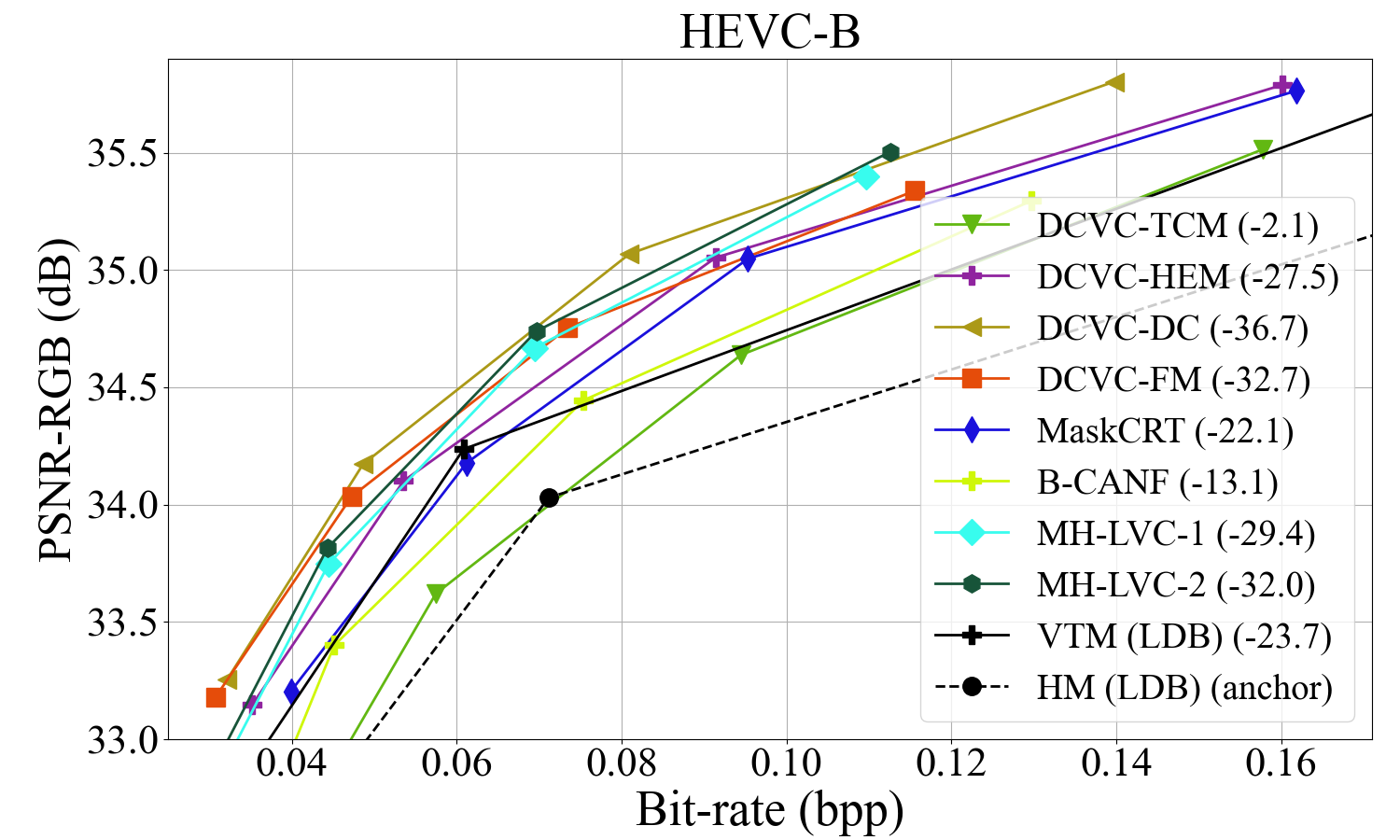}
    \end{subfigure}
    \begin{subfigure}{0.45\linewidth}
        \centering
        \includegraphics[width=\linewidth]{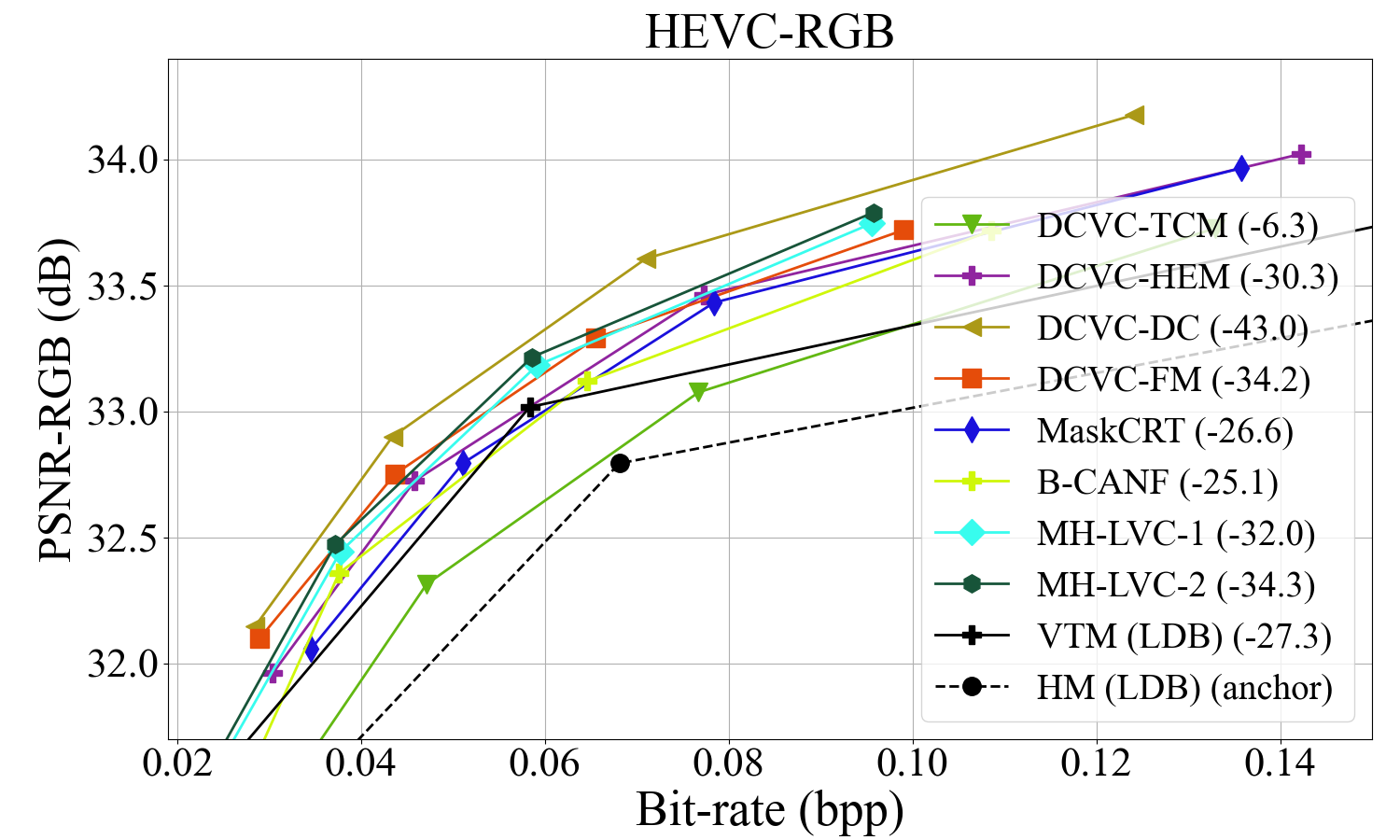}
    \end{subfigure}
    \begin{subfigure}{0.45\linewidth}
        \centering
        \includegraphics[width=\linewidth]{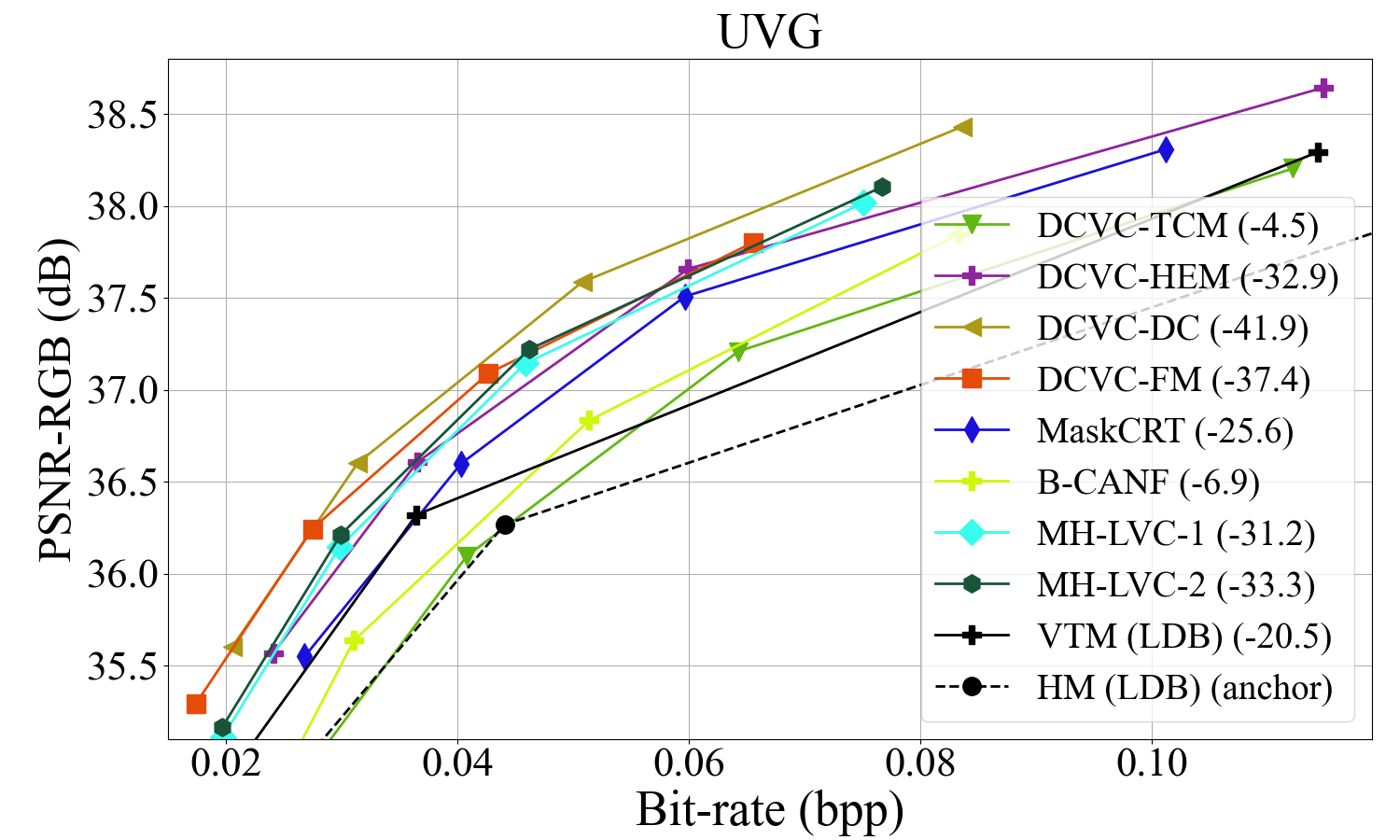}
    \end{subfigure}
    \begin{subfigure}{0.45\linewidth}
        \centering
        \includegraphics[width=\linewidth]{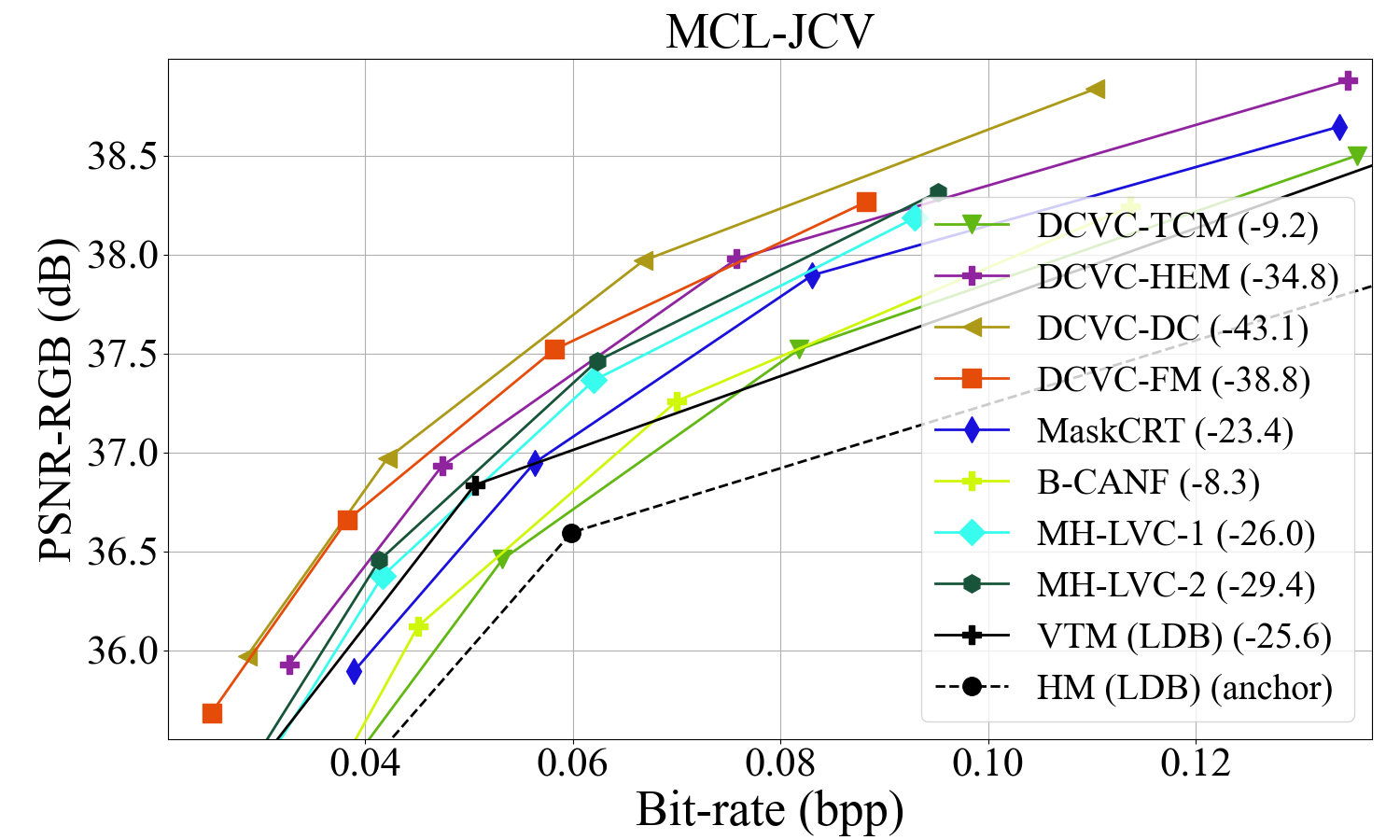}
    \end{subfigure}
    \caption{Rate-distortion performance comparison with intra-period 32 in terms of PSNR-RGB. The numbers within the parentheses are BD-rates, with HM-16.25 (Low-delay B) serving as the anchor.}
    \label{fig:supp_RGB_I32}
    \end{center}
\end{figure*}

%% file: table/Supp_RD_Plot/RGB_INF.tex
\begin{figure*}[tbp]
    \begin{center}
    \begin{subfigure}{0.45\linewidth}
        \centering
        \includegraphics[width=\linewidth]{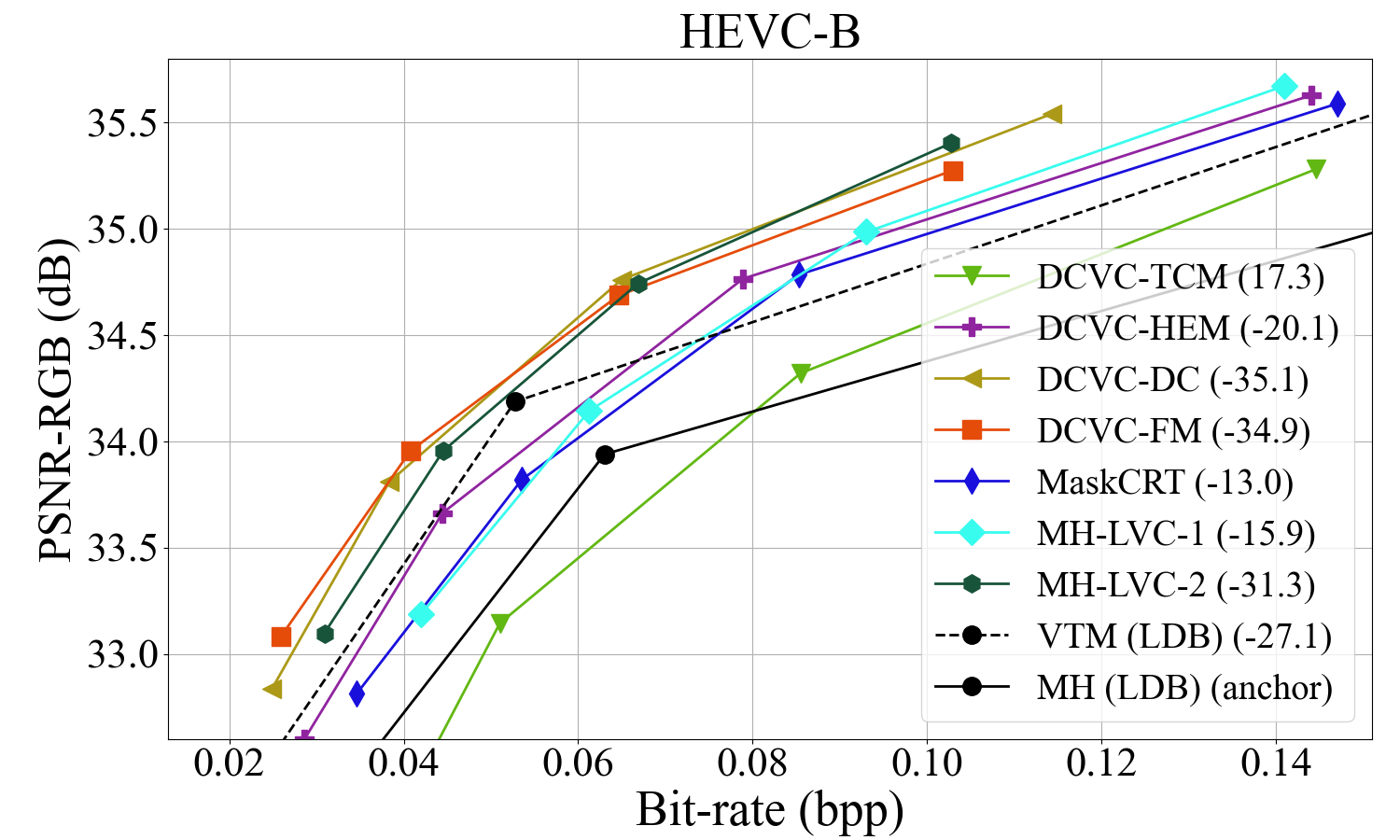}
    \end{subfigure}
    \begin{subfigure}{0.45\linewidth}
        \centering
        \includegraphics[width=\linewidth]{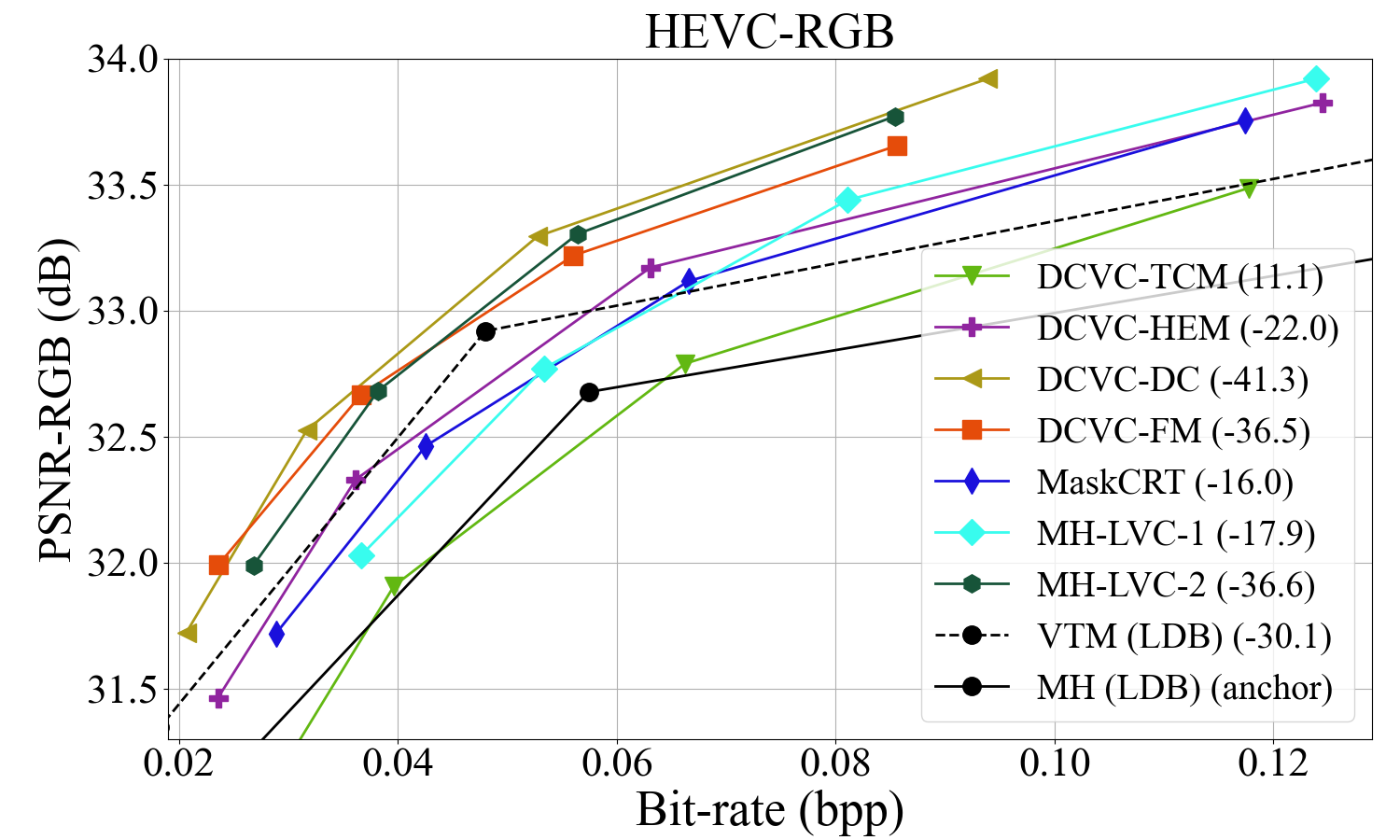}
    \end{subfigure}
    \begin{subfigure}{0.45\linewidth}
        \centering
        \includegraphics[width=\linewidth]{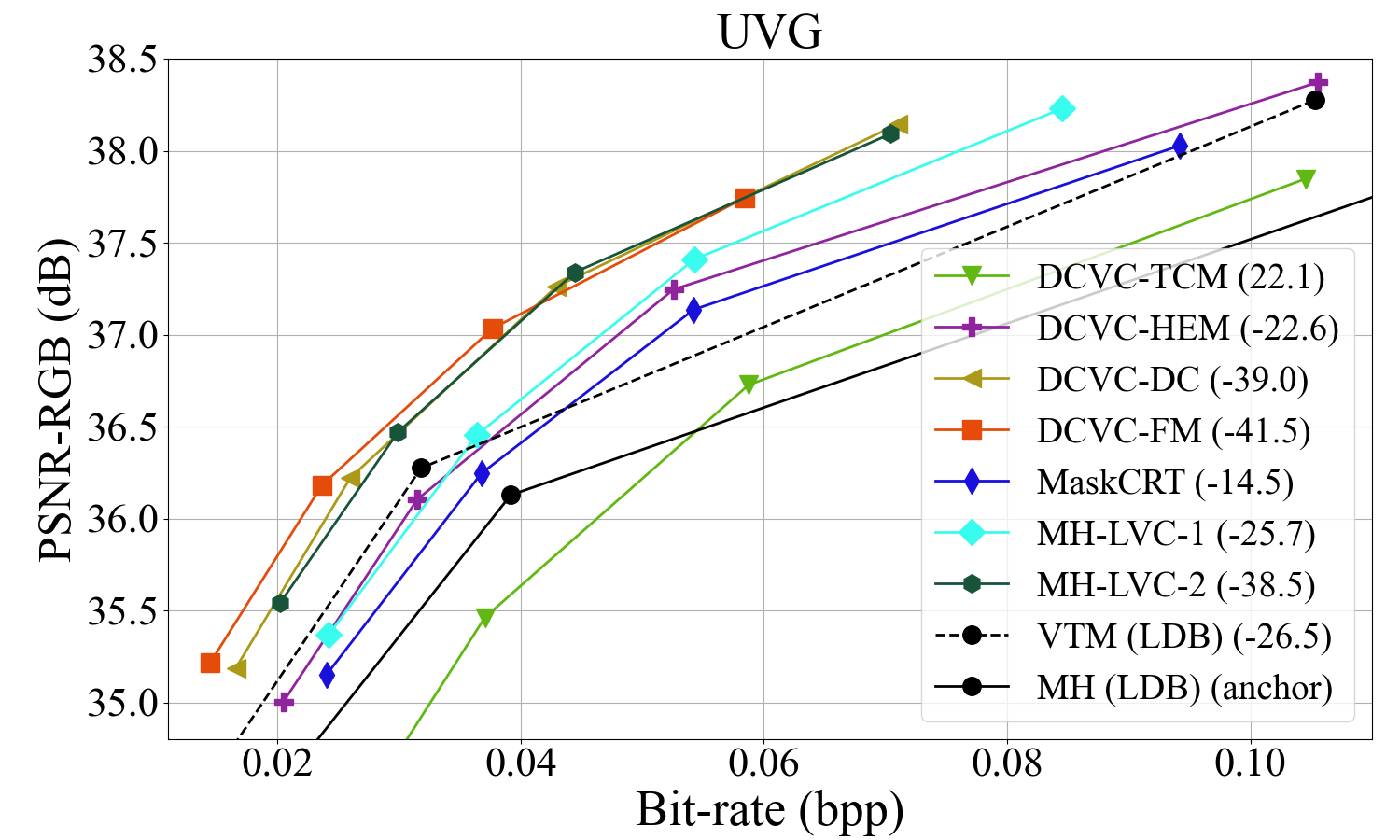}
    \end{subfigure}
    \begin{subfigure}{0.45\linewidth}
        \centering
        \includegraphics[width=\linewidth]{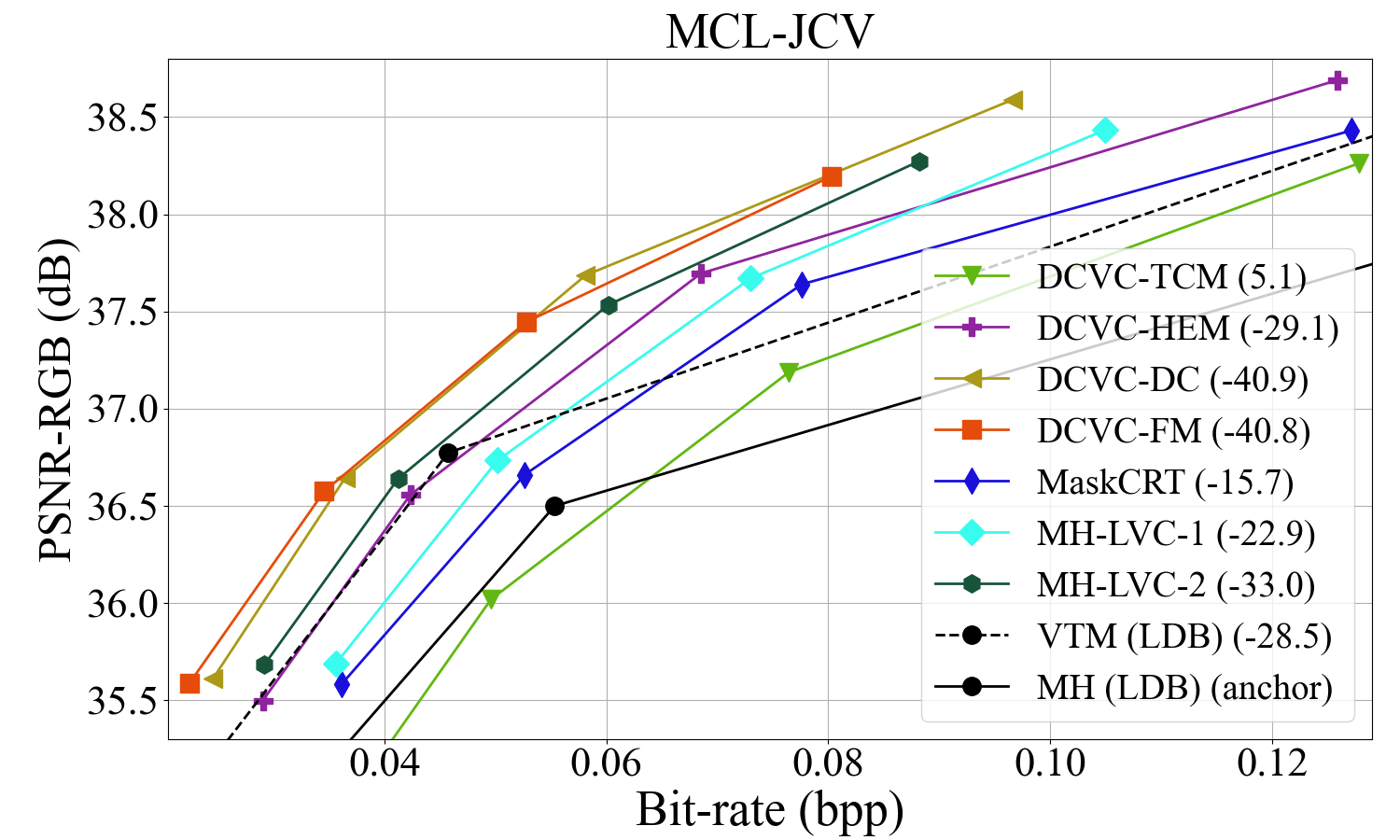}
    \end{subfigure}
    \caption{Rate-distortion performance comparison under an infinite intra-period in terms of PSNR-RGB. The numbers within the parentheses are BD-rates, with HM-16.25 (Low-delay B) serving as the anchor.}
    \label{fig:supp_RGB_INF}
    \end{center}
\end{figure*}

%% file: section/K_More_Visualization.tex
\input{section/F-2_Mask_Visualization}

\section{More Visualizations}
\label{sec:more_vis}
 Fig. \ref{fig:supp_mask_vis} presents more visualizations for the gating signal generated by the spatial gate predictor.

%% file: section/F-2_Mask_Visualization.tex
\begin{figure*}[tbp]
\centering
\footnotesize
{
    \includegraphics[width=0.95\linewidth]{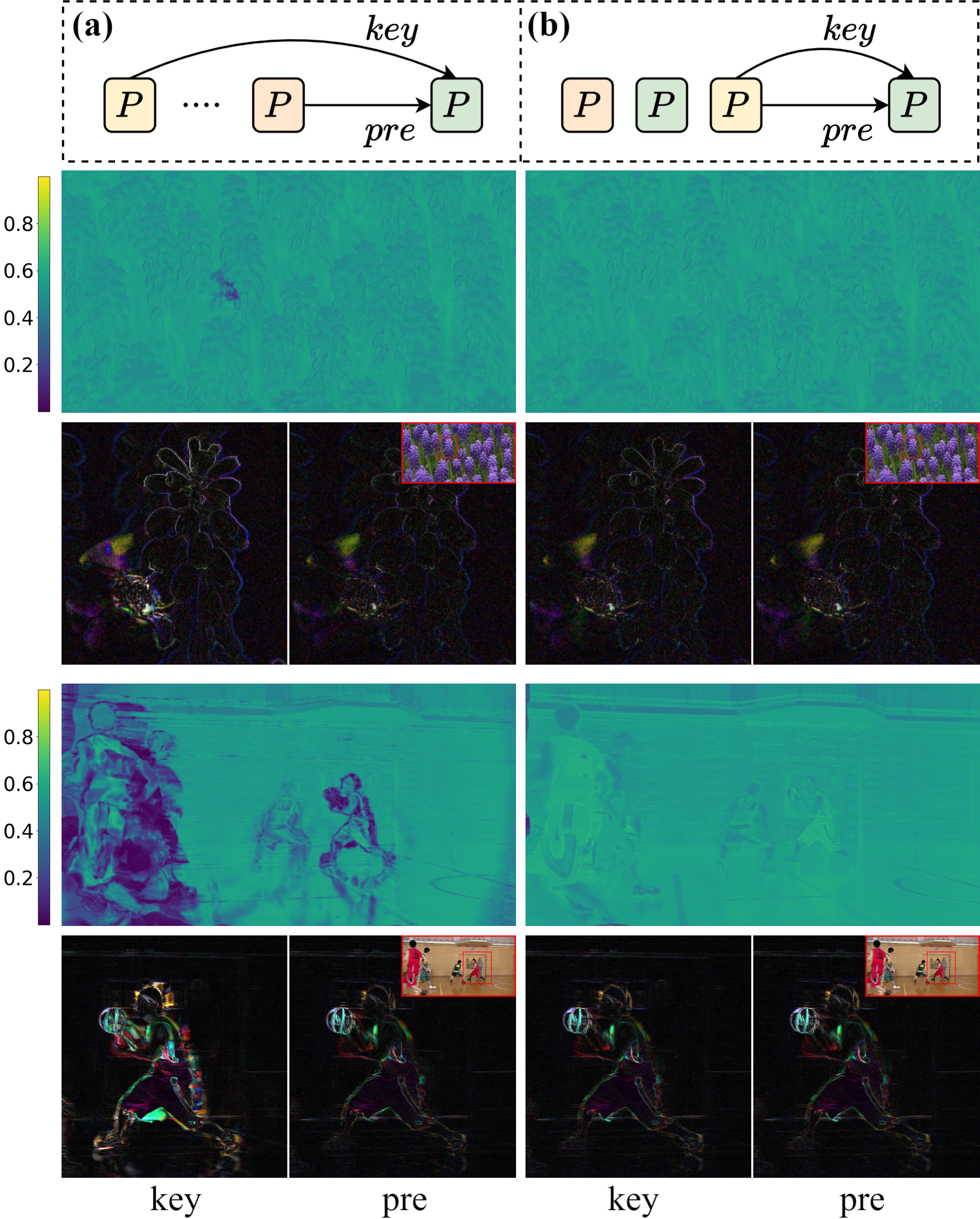} 
    \\
    
}

\vspace{-1mm}

\caption{Visualization of the gating signal $\gamma^{(1)}$ for two temporal prediction structures. (a) adopts both long- and short-term reference frames, (b) has two predictors derived from the same short-term reference frame with the same optical flow map. The bottom row displays the prediction residues between the coding frame $x_t$ and its two motion-compensated reference frames $\hat{x}_{key}$ (denoted as key) and  $\hat{x}_{t-1}$ (denoted as pre).}

\label{fig:supp_mask_vis}

\end{figure*}